\documentstyle{mn2e}

\title[Aligned Radio Polarizations?] {The polarization in the
JVAS/CLASS flat-spectrum radio sources: II. A
search for aligned radio polarizations \\} 
\author[Joshi et al]
{S.A. Joshi, R.A. Battye, I.W.A. Browne, N. Jackson, T.W.B. Muxlow, and
\newauthor P.N. Wilkinson\\
The University of Manchester, Jodrell Bank Observatory, Macclesfield,
Cheshire, SK11 9DL, U.K.\\}

\begin{document}
\input{psfig}
\label{firstpage}
\maketitle

\begin{abstract}

We have used the very large JVAS/CLASS 8.4-GHz surveys of
flat-spectrum radio sources to test the hypothesis that there is a
systematic alignment of polarization position angle vectors on cosmological
scales of the type claimed by Hutsem\'ekers et al. (2005). The
polarization position angles of 4290 sources with polarized flux
density $\geq$1~mJy have been examined. They do not reveal large-scale
alignments either as a whole or when split in half into high-redshift 
(z $\ge$ 1.24) and low-redshift sub-samples. 
Nor do the radio sources which lie in the
specific areas covered by Hutsem\'ekers et al. (2005) show any
significant effect. We have also looked at the position angles of
parsec-scale jets derived from VLBI observations and again find no
evidence for systematic alignments. Finally, we have investigated the
correlation between the polarization position angle and those of the
parsec-scale jets. As expected, we find that there is a tendency for
the polarization angles to be perpendicular to the jet
angles. However, the difference in jet and polarization position
angles does not show any systematic trend in different parts of the
sky. 

\end{abstract}


\section{Introduction}

In an isotropic universe the sizes of the largest coherent structures
are expected to be limited by the time taken for the primordial
density fluctuations to undergo gravitational collapse. In practice
both observations of large-scale structures and the latest N-body
simulations agree that the  
largest structures have scales of $\leq$100~Mpc. It is therefore
obvious that any observation of coherent behaviour of objects on much
larger scales than $\sim$100~Mpc would have profound implications for
our understanding of the nature of the Universe. Over the years there
have in fact been several reports of the detection of large-scale
coherence, mostly based on measurements of the linear polarizations of
active galaxies, and more recently based on observations of the cosmic
microwave background (CMB). 

Birch (1982) claimed to have found evidence for a rotating universe by
comparing the radio polarization position angles for extended radio
sources with the direction of elongation of radio sources.  The claim
was that there were systematic offsets between the two position angles
and that these were systematically different in widely different part
of the sky. Though attacked on statistical grounds at the time
(e.g. Phinney \& Webster 1983) it is not clear that this ``Birch
effect'' has totally gone away (Kendall \& Young 1984; Jain et al.,
2003). An independent effect claimed by Nodland and Ralston (1997) was
that there is a systematic rotation of plane of polarization over
cosmological distances. But Wardle et al (1997) refute their claim by
pointing out that in individual radio sources the polarization
position angles are strongly correlated with extended radio
morphology, even at high redshift which excludes large cosmological
rotations. Also the Nodland and Ralston claim has been attacked on the
basis of statistics (e.g. Carroll and Field (1997), Eisenstien and
Bunn (1997), Loredo et al (1997)).

More recently, Hutsem\'ekers (1998), Hutsem\'ekers \& Lamy (2001) and
Hutsem\'ekers et al. (2005) have been accumulating evidence that the
linear polarizations of quasars in optical wavelengths are
non-uniformly distributed, being systematically different near the
North and South Galactic poles. They also find that the clustering of
position angles is more evident if they divide their sample of 355
sources by redshift. They use this fact to argue against the effect
being a result of the polarization of the quasar light as it passes
through Galactic interstellar dust since this would affect all
redshifts indiscriminately. 

Stimulated by these hints of interesting polarization effects, and
reports of a preferred axis in CMB fluctuations (Schwarz et al., 2004;
Land \& Magueijo, 2005; Raeth et al., 2007), we decided to examine
some linear polarization measurements we had made with the Very Large
Array (VLA) between 1990 and 1999 of a large sample of compact
flat-spectrum radio sources. The JVAS (Jodrell Bank-VLA Astrometric
Survey) programme from 1990-1992 (Patnaik et al. 1992, Browne et
al. 1998, Wilkinson et al. 1998) was undertaken with the two main
objectives of measuring accurate positions for sources to find
interferometer phase calibrators and to identify gravitationally
lensed sources. However, we also obtained data for the linear
polarizations of the 2308 sources. In addition, the later CLASS survey
(1994-1999; Myers et al. 2003, Browne et al. 2003) targeted fainter
compact radio sources and the majority of these observations have
suitable calibrators available to derive degrees of polarization. The
total size of the JVAS/CLASS surveys is 16503 sources, and as we
discuss in Jackson et al., (2007) (hereafter Paper I), 4290 have
polarized flux density $\ge$1 mJy. These 4290 sources we make use of
for studies of possible polarization alignments. 

\begin{figure*}{\label{JCaitoff}}
  \psfig{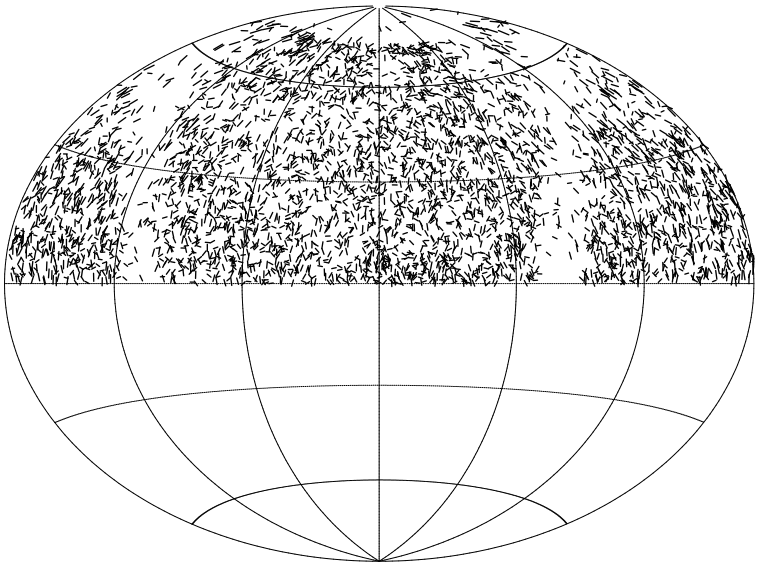}
  \caption{Aitoff plot for all the objects in the sample we used, from
  JVAS/CLASS samples. The lack of uniformity arises from the exclusion
  of low Galactic latitude sources from CLASS and the fact that in some
  CLASS regions we were not able to calibrate the polarizations to our
  satisfaction.}
\end{figure*}

On its way to the observer polarized radiation passes both through the
host galaxy of the AGN and the Galaxy and suffers Faraday
rotation. Since it is dependent on the square of the wavelength,
Faraday rotation is therefore a much more significant effect at radio
than at optical wavelengths. However, as we have shown in Paper I,
total rotation measures are generally much too small to destroy
information about the intrinsic position angles of polarization when
the measurements are made at 8.4~GHz.

\section{Initial Results and Reanalysis}{\label{initial_arp}}

Our initial examination of JVAS for evidence of aligned radio
polarizations made use of the original calibrations of the data done
primarily with astrometry in mind and not the accurate measurement of
source polarizations. Using these data we did initially find apparent
systematic alignments so we decided on a careful re-calibration of
both the JVAS and CLASS data, concentrating on polarizations. The
results of this recalibration and analysis are presented in Paper I.

\subsection{The data and analysis}

We make use of the 4290 sources presented in Paper I and full details
of the calibration and analysis are 
given therein. See Fig. {\ref{JCaitoff}} for the Aitoff projection of
the data. Here we only briefly address calibration and analysis issues 
which may be directly relevant to any search for systematic alignments
of polarization position angles within the sample. These issues are:

\begin{itemize}

\item {\it Polarization residual calibration.} Incorrect removal of
instrumental polarization residuals can lead to a bias in the measured
polarization angles. The JVAS/CLASS observations were concentrated in
different regions of sky in different observing runs and thus it is
possible that poor, and epoch-dependent, instrumental residual removal
could lead to 
apparent large scale alignments in different regions of sky. On the
other hand, incorrect residual removal is very unlikely to mask real
regions of aligned position angles as the errors would have to
conspire in such a way as to randomize the observed position angles in
regions of true alignment.

The analysis presented in Paper I leads us to believe that systematic
errors in the residual calibrations are at a level $\sim 0.3\%$.

\item {\it Polarization position angle calibration.} Errors in
position-angle calibration would lead to position angles being
systematically wrong, perhaps by different amounts in different
observing runs.  We believe that errors in the position angle
calibration are at the level of $\leq 10^{\circ}$. Position-angle
calibration errors cannot produce areas of systematic alignments where
none exist. They 
could, however, reduce the prominence of areas of real alignment if
the errors were large enough and changed systematically across a
region of real alignment. But at $\leq 10^{\circ}$ these errors are small
and, in fact, comparable to the random errors on the majority of
sources. Thus they are not a good reason for missing real alignments.

\item {\it Data analysis -- CLEAN bias.} Even with perfect calibration
it is possible for the analysis technique adopted to lead to
systematic biases in the distribution of polarization position
angles. In particular the CLEAN deconvolution process suffers from a
tendency to bias flux densities towards zero by an amount that depends
on the number of CLEAN iterations. For total intensity data this is
relatively benign, but for the analysis of polarization data where
Stokes Q and U maps are separately cleaned it can cause significant
problems when Q and U are combined to give a position angle. This is
because Q and U can be negative and because polarized flux densities
are generally low. Hence biases towards zero are proportionately more
significant and result in a disproportionate number of sources having
apparent position angles around zero, $\pm45^{\circ}$ and
$\pm90^{\circ}$. The issue of CLEAN bias and other systematic effects 
which may affect polarization measurements are discussed more extensively
in Battye et al. (in prep). It is to avoid the effects of CLEAN
bias that the results we analyse in this paper have all been obtained by
model fitting in the visibility plane without the use of CLEAN.

\end{itemize}

\subsection{Westerbork Data at 5 GHz}{\label{WSRT-text}}

JVAS and CLASS were made using the VLA which has feeds that produce
opposite hands of circular polarization that can then be
cross-correlated to produce linear polarizations (Stokes Q and U). We
thought it useful to check the accuracy of the VLA polarization
results by comparing them with measurements obtained in a different
manner. Also, if the measurements were done at a different frequency
this would give some Faraday rotation information.(See Section
{\ref{discussion}}) We therefore observed 340 sources from JVAS/CLASS
lists with Westerbork Synthesis Radio Array (WSRT) at 5~GHz. This uses
linear-polarization feeds to obtain the polarization information; thus
the methodology is completely different from that of the VLA. 

All the sources observed were within 0 $\le$ $\delta$ $\le$
40$^{\circ}$, and 
had polarized flux density, $P\ge$ 4 mJy at 8.4~GHz. Not all 
the measurements were successful and we obtained useful polarization
data for 336 objects. Fig. 2 compares position angles
(PA) from JVAS/CLASS and WSRT data; the difference clearly peaks at
0$^{\circ}$. 

Full details of the observations, their calibration and analysis will
be given elsewhere (Joshi et al, in prep). Briefly, CTD 93 and 3C 286
were observed as flux and polarization calibrators. The data analysis
was performed using the NEWSTAR package. The target sources were
assumed to be point-like and the values for Stokes $Q$, $U$ (and V)
were obtained by model fitting.

\begin{figure*}{\label{figJC-WSRT}}
  \psfig{figure=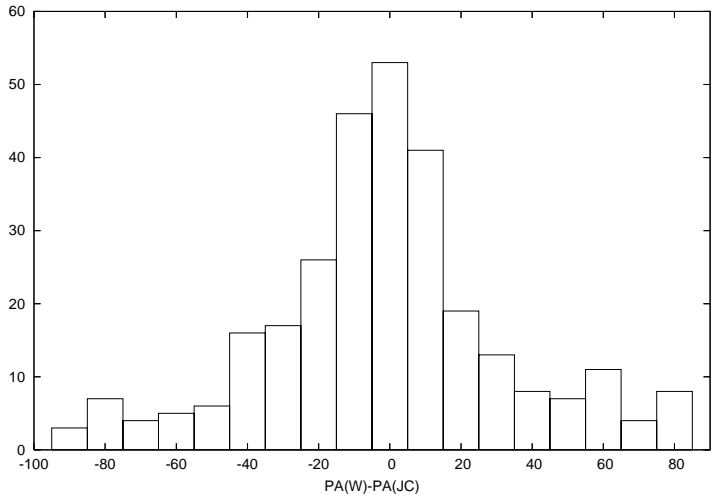,width=10cm,angle=270}
  \caption{The histogram of the differences between position angles
  measured using the WSRT and the VLA in the JVAS/CLASS surveys.}
\end{figure*}

\section{Statistical analysis}{\label{stat}}

Our polarization data as described in Paper I have
been analyzed for signs of statistical non-uniformity of the type
claimed by Hutsem\'ekers et al (2005). In Paper I it was shown that
the distribution of position angles for the sample treated as a whole
was consistent with uniformity. However, this does not preclude the
possibility of sub-areas in which there are locally
strong alignments. Four tests have been performed. In the first two
the sky has been divided into ``tiles'' defined using the HEALPix
software package (G\'orski et al., 2005) and the distribution within
each tile has been tested for non-uniformity using a $\chi^2$ test on
the histograms and by adding the polarization vectors to form a random
walk. In the third test a search has been made for clusters of aligned
position angles using a nearest neighbour analysis. In the last test
we implemented one of the approaches used by Hutsem\'ekers et al. All
these tests applied to the 4290 objects from JVAS and CLASS catalogues
for which $P\ge$ 1mJy.

\subsection{Histogram method}{\label{localhist}}

We have investigated a simple adaptation of the histogram test
performed on whole dataset in Paper I which is designed to search for
localized non-uniform behaviour. The data were split up into pixels
defined by the HEALPix software and the $\chi^2$ of each of these
histograms of position angles was computed on the assumption that the
expected distribution was uniform. This value is then compared to the
$\chi^2$ found for 10000 realizations of sources at the same
positions, but with PAs which are drawn from a uniform
distribution. For the real data the number of pixels that have a
$\chi^2$ indicating that the distribution has failed the test for
uniformity at different degrees of significance is noted. This
is compared to the number one might expect for that number of pixels

The size of each pixel is defined by the HEALPix variable $n_{\rm
  side}$, which gives the number of pixels in the whole sky $n_{\rm
  pix}=12n_{\rm side}^2$, and hence the area of a pixel is
  $\Omega_{\rm pix}=\theta_{\rm pix}^2=4\pi/n_{\rm pix}$. We also need
  to choose a bin size for the histogram which is defined by $n_{\rm
  bin}$, the number of bins covering the range $-90^{\circ}$ to
  $90^{\circ}$, which is varied in the analysis. Moreover, we have
  performed this test using two different coordinate systems
  (celestial and galactic) so as to investigate the effects of choice
  of pixel shape defined by HEALPix.

\begin{table}
\begin{center}
\begin{tabular}{|cc|cc|cc|}
\hline
 $\theta_{\rm pix}/{\rm deg}$ & $n_{\rm obj}$ & $N(95\%)$ & $P(95\%)$ & $N(99\%)$ & $P(99\%)$ \\
\hline
58.5 & 8 & 1 & 12.5 & 0 & 0 \\
29.3 & 28 & 2 & 7.1 & 1 & 3.5 \\
14.7 & 103 & 3 & 2.9 & 2 & 1.9 \\
7.3 & 392 & 10 & 2.5 & 5 & 1.2 \\
3.7 & 1342 & 7 & 0.5 & 1 & 0.07 \\
\hline
\end{tabular}
\end{center}
\caption{Results of the histogram test. $\theta_{\rm pix}$ is the
  approximate pixel size, $n_{\rm obj}$ is the number of pixels
  containing an object. $N(95\%)$ and $N(99\%)$ are the number of
  pixels which fail the test at 95\% and 99\% respectively, that is,
  those pixels for which the $\chi^2$ is greater than that found in 500 and
  100 of the random 10000 realizations, respectively. $P(95\%)$ and $P(99\%)$
  quantify the number of failures as a percentage of the pixels
  containing a source. }
\label{tab:hist} 
\end{table}

The results of this test are presented in Table~\ref{tab:hist} for
$n_{\rm bin}=4$ and the celestial coordinate system. For large pixel
sizes ($\theta_{\rm pix}\approx 58.5^{\circ}$ and $29.3^{\circ}$)
there is a slight excess of pixels which fail the test - one would
expect around $5\%$ of pixels to fail the test at $95\%$ confidence,
for example - and substantially less failing than one expects for
$\theta_{\rm pix}\approx 3.7^{\circ}$. The latter result can be
understood by realizing that there are, on average, less than 5
sources per pixel in this case. The results for large pixel sizes can
be ``explained'' as being due to small number statistics. For $\theta_{\rm
  pix}=14.7^{\circ}$ and $7.3^{\circ}$, things seem compatible with
there being no excess alignment, over and above that which is
expected. 

This test is however not completely robust since it relies on a choice
of $n_{\rm bin}$ and the coordinate system. We have checked that the 
results presented in Table~\ref{tab:hist} are not particularly
sensitive to these choices by using $n_{\rm bin}=6$, and also a
galactic based coordinate system. Hence, we conclude that the test is
compatible with the null hypothesis that the PAs are chosen from a
uniform distribution. 

\subsection{Random walk test}{\label{randwalk}}

We have performed another test using the HEALPix-defined pixelation
scheme: for each source in the pixel we have constructed a unit vector
in the direction of the polarization ($\hat p_{\rm i}=(\sin
2\theta,\cos 2\theta)$) and these are then added together to form a
random walk. Again the results are compared to 10000 realizations
whose PAs are drawn from a uniform distribution. In contrast to the
histogram test, this does not require the choice of $n_{\rm bin}$; it
also seems to work better when there are smaller numbers of sources in
the pixels. It does, however, still depend on a choice of coordinate
system. 

The results of this test are presented in Table~\ref{tab:rand}. We see
that again that there is a slight excess of pixels which fail the test
for large pixel sizes, but that the number failing the test appears
compatible with uniformity for lower values of $\theta_{\rm pix}$.

\begin{table}
\begin{center}
\begin{tabular}{|cc|cc|cc|}
\hline
 $\theta_{\rm pix}/{\rm deg}$ & $n_{\rm obj}$ & $N(95\%)$ & $P(95\%)$ & $N(99\%)$ & $P(99\%)$ \\
\hline
58.5 & 8 & 1 & 12.5 & 0 & 0 \\
29.3 & 28 & 3 & 10.1 & 1 & 3.5 \\
14.7 & 103 & 2 & 1.9 & 0 & 0 \\
7.3 & 392 & 16 & 4.0 & 4 & 1.0 \\
3.7 & 1342 & 56 & 4.1 & 7 & 0.5 \\
\hline
\end{tabular}
\end{center}
\caption{Results from the random walk test. The columns are the same
as in Table~\ref{tab:hist}}
\label{tab:rand}
\end{table}

\subsection{Nearest Neighbour Method} {\label{neighbour_method}}

As a simple global test to see if the PA are aligned we used the
following method: for every object, the $n$ nearest objects are
checked for alignment. If the object has PA within $\Delta\theta$ of
the object under consideration, it is 
considered aligned. The choice of the appropriate value of $n$ depends
on the size of the sample, since the probability of finding $n$
alignments by chance should be low. It also depends upon the scale of
any real clustering in the data which is, of course, unknown. However,
if the regions over which alignments occur contain more than $n$
objects, the test will pick out many objects in the same area all of
which will have significantly more rear neighbours having aligned
polarizations than expected by chance and real structures will not be
missed. We initially choose a value of $n$ = 25 but have explored
other values up to a maximum of $n$ = 250. We typically used
$\Delta\theta=45^{\circ}$ since all the errors on the PA measurements
are significantly smaller than this angle.

\begin{figure*}
\begin{tabular}{lr}
  \psfig{figure=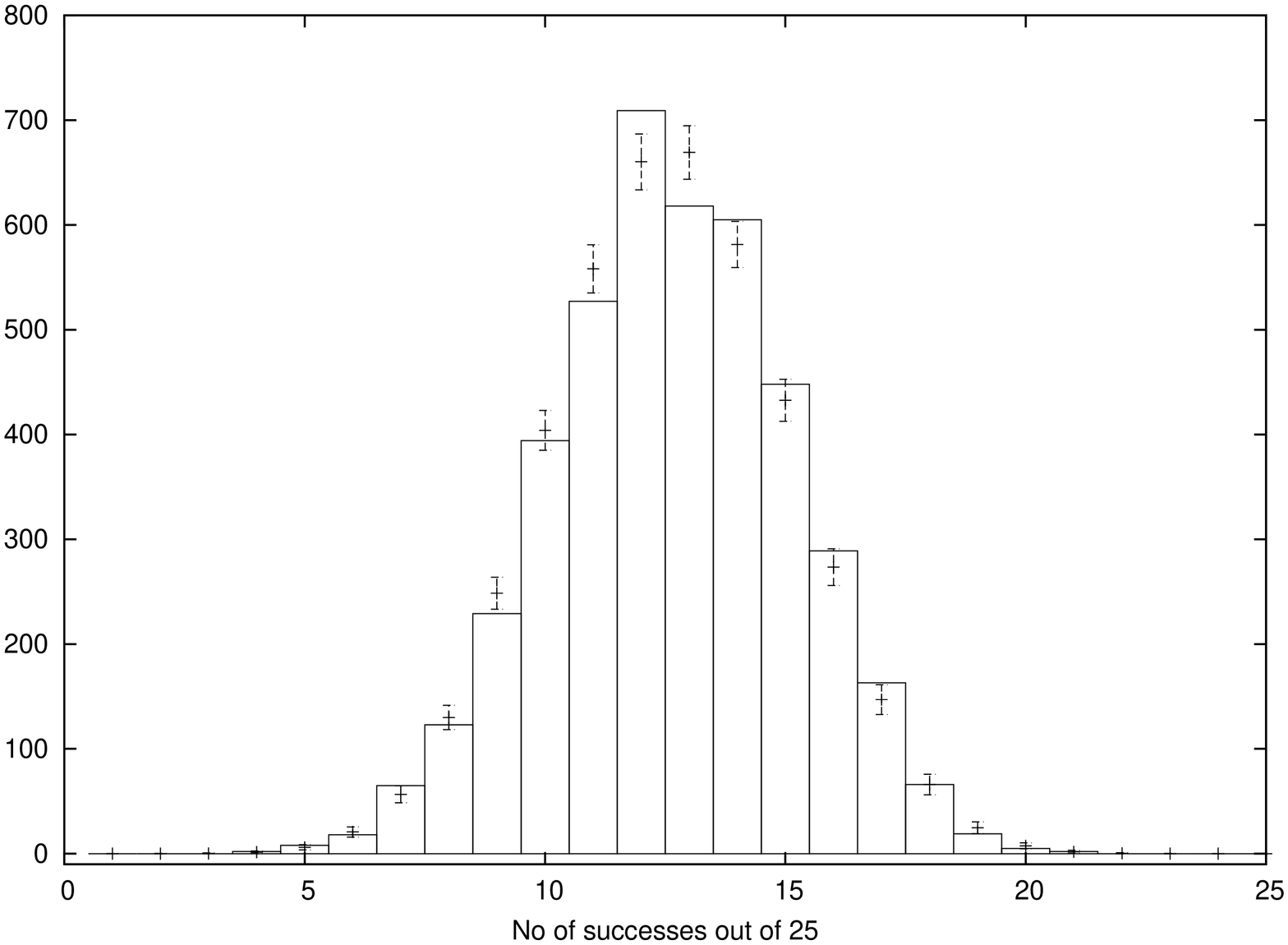,width=8cm,angle=270}
  \psfig{figure=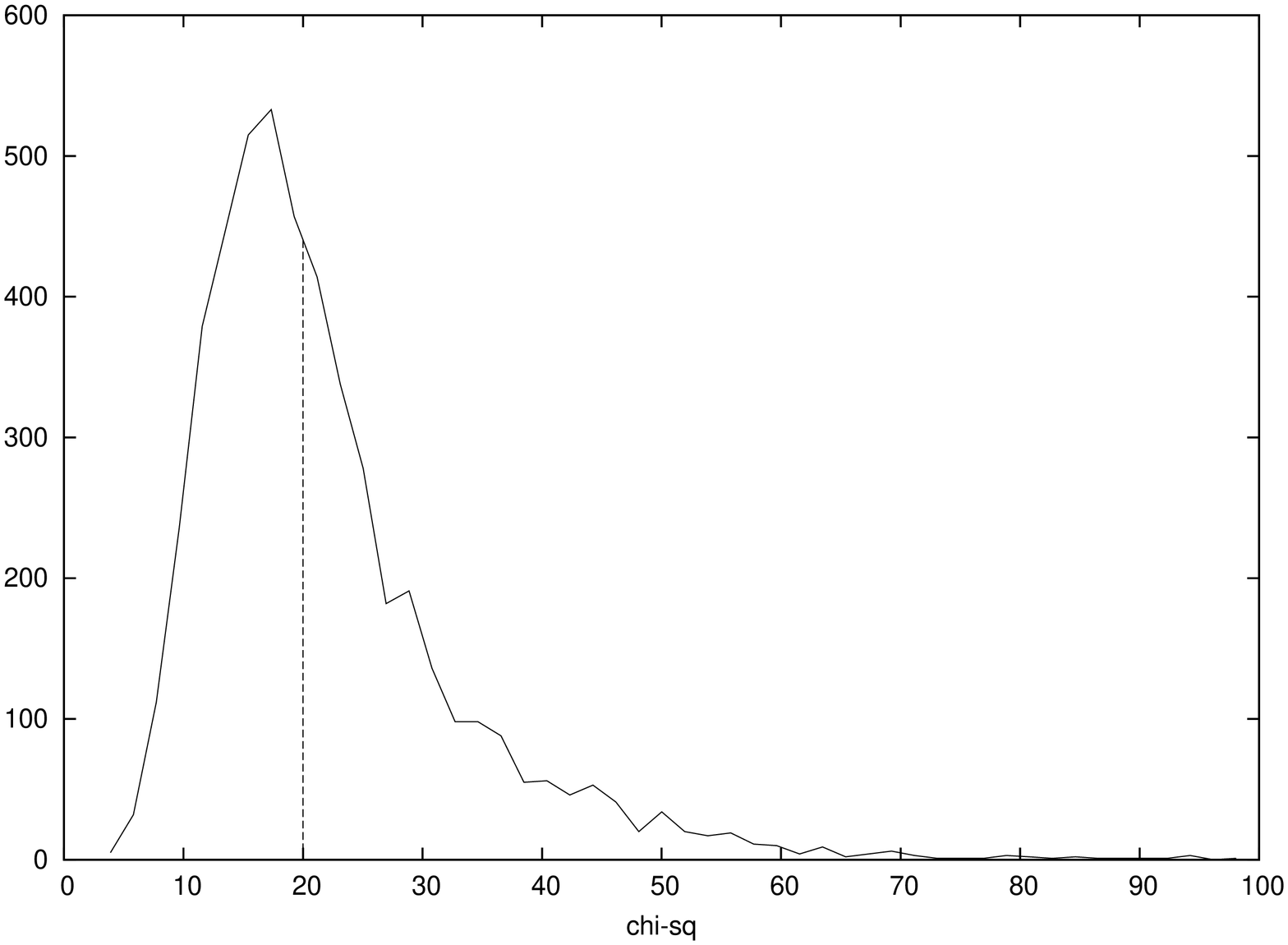,width=8cm,angle=270}
\end{tabular}
  \caption{Comparison of nearest neighbour distribution for the 4290
    JVAS/CLASS objects with that expected for a randomly oriented
    sample obtained from 5000 random realizations. The solid
    lined-histogram is for the real data and dashed data with error
    bars is from 5000 random samples. $\chi^2$ for this distribution
    is 20.1. The right hand panel shows the distribution of $\chi^2$
    obtained by comparing each random realization with the averaged of
    all the random ones. The line represents $\chi^2$ for the observed
    sky.}
\label{fig:JCwhole}
\end{figure*}

For the whole sample a histogram showing the number of objects having
N out of 25 aligned nearest neighbours is plotted in Fig
{\ref{fig:JCwhole}}. The error 
bars have been derived from 5000 random realizations where, keeping
positions of the objects the same, the sky was populated with sources
having random PAs and the nearest neighbour test performed on each
realization thus producing 5000 histograms like the one shown in Fig
{\ref{fig:JCwhole}}. For each bin in the histograms the distribution
of values was obtained 
and fitted by a Gaussian. The plotted points and error bars are,
respectively, the means and standard deviations obtained from the
Gaussian fits.
It is clear that, given the error bars, the real histogram is a
reasonably good fit to that expected for a sample with randomly
orientated position angles; the $\chi^2$ for the real distribution
when compared with the average random distribution is 20.1. The
appropriate number of degrees of freedom is not obvious because 
the numbers in the histogram bins are not statistically independent.
In Fig. {\ref{fig:JCwhole}}b we show the distribution of $\chi^2$,
obtained from the random realizations, by comparing each with the mean
of the 5000, with the line marking the $\chi^2$ for the real sky. The
probability of finding a $\chi^2$ equal to or greater than the
observed one is 45\%. 

\subsection{Hutsem\'ekers' test}

\begin{figure*}
\psfig{figure=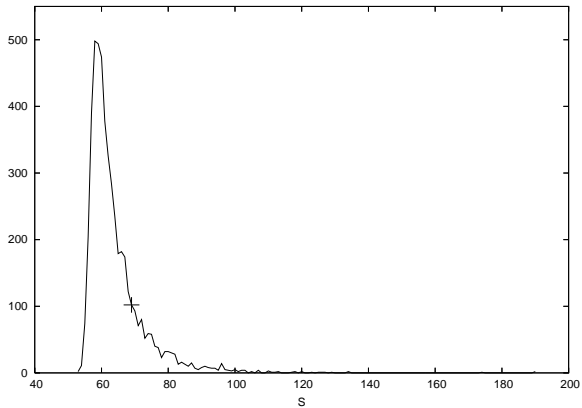,width=8cm,angle=0}
\caption{The distribution of the parameter S for 5000 random samples. For
the observed data, S is 68.6 and is marked by a line in the figure.}
\label{S_distribution}
\end{figure*}

Hutsem\'ekers (1998) used a dedicated statistical test to identify if
the polarization vectors are aligned in 2D or 3D space. It is more
sophisticated than our own nearest neighbour test in that it gives
extra weight to aligned objects according to their degree of alignment
and distance apart. (See Hutsem\'ekers (1998), Section 5.1 for the
complete description.) As in our implementation of our nearest
neighbour test, alignments were searched for within 
$\sim$45$^\circ$ in groups of 25 objects. We have implemented the
Hutsem\'ekers (1998) test on JVAS/CLASS 
data in 2D (since we have incomplete redshift information).  In this
test, a parameter, S is calculated which depends upon the spatial
distribution of objects, their PAs and the distribution of these. We
find S = 68.6 for the 4290 sources in the JVAS/CLASS sample. To test
whether or not this value of S indicates statistically significant
clustering we generated 5000 random samples. For each sample the positions of
the objects are same as those observed but PAs are randomized. The
S distribution obtained from the random samples is shown in figure
{\ref{S_distribution}}. Using this distribution we find that the
probability of finding S $\geq$ 68.6 occurring by chance is 19\%. 

\section{Division of the sample into redshift bins}

\begin{figure*}
 \begin{tabular}{lr}
   \psfig{figure=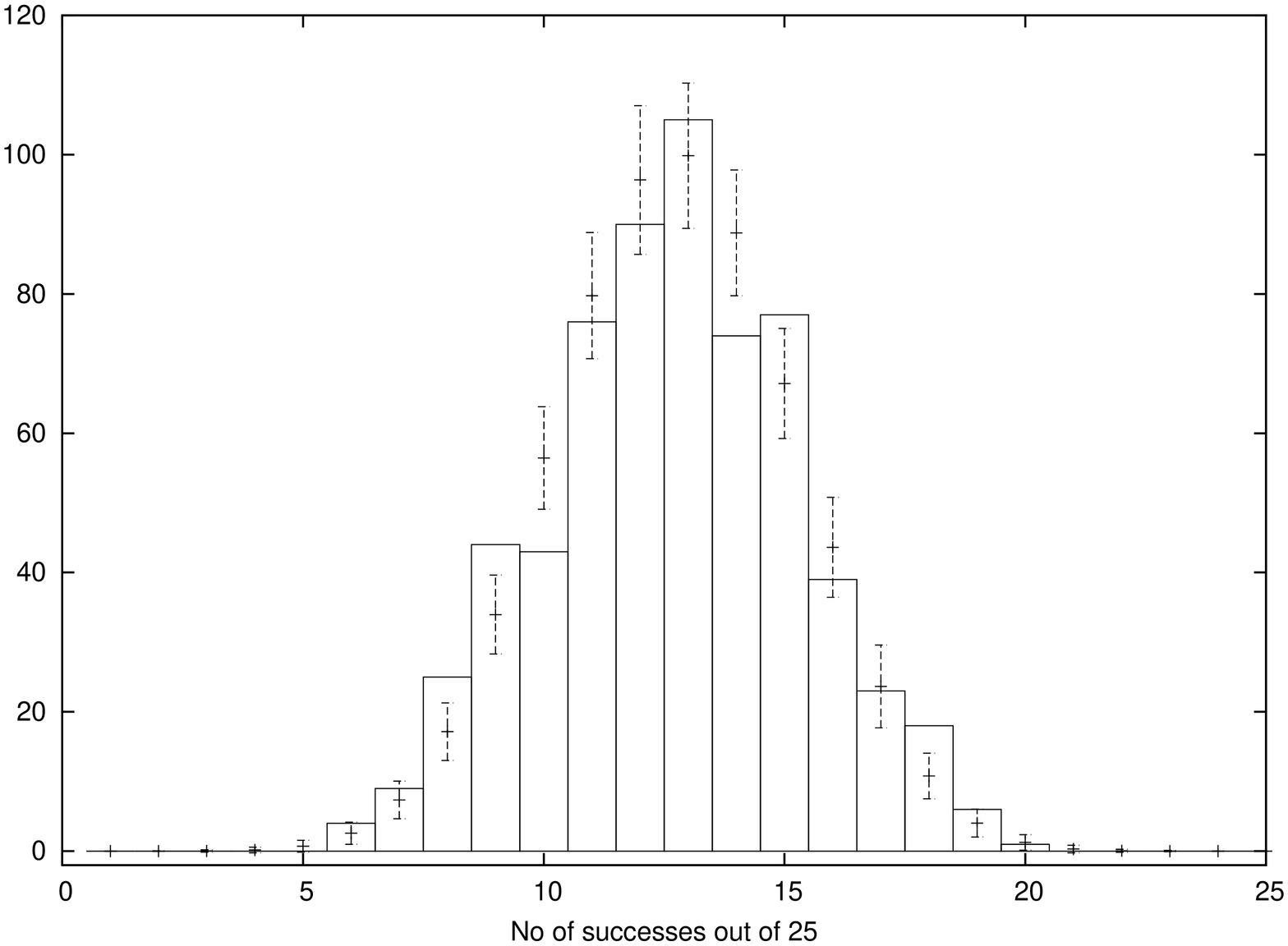,width=8cm,angle=270}
   \psfig{figure=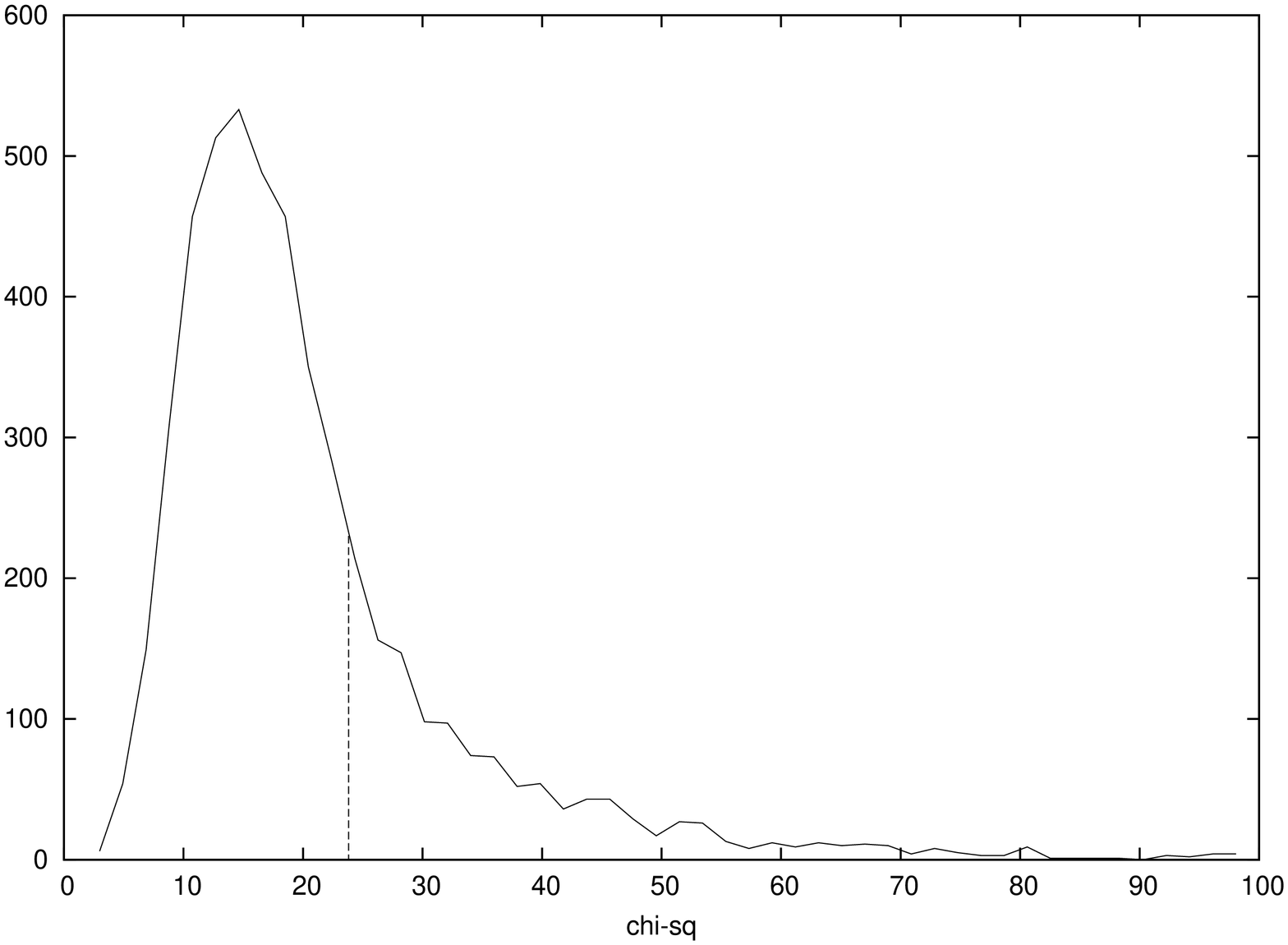,width=8cm,angle=270}\\
   \psfig{figure=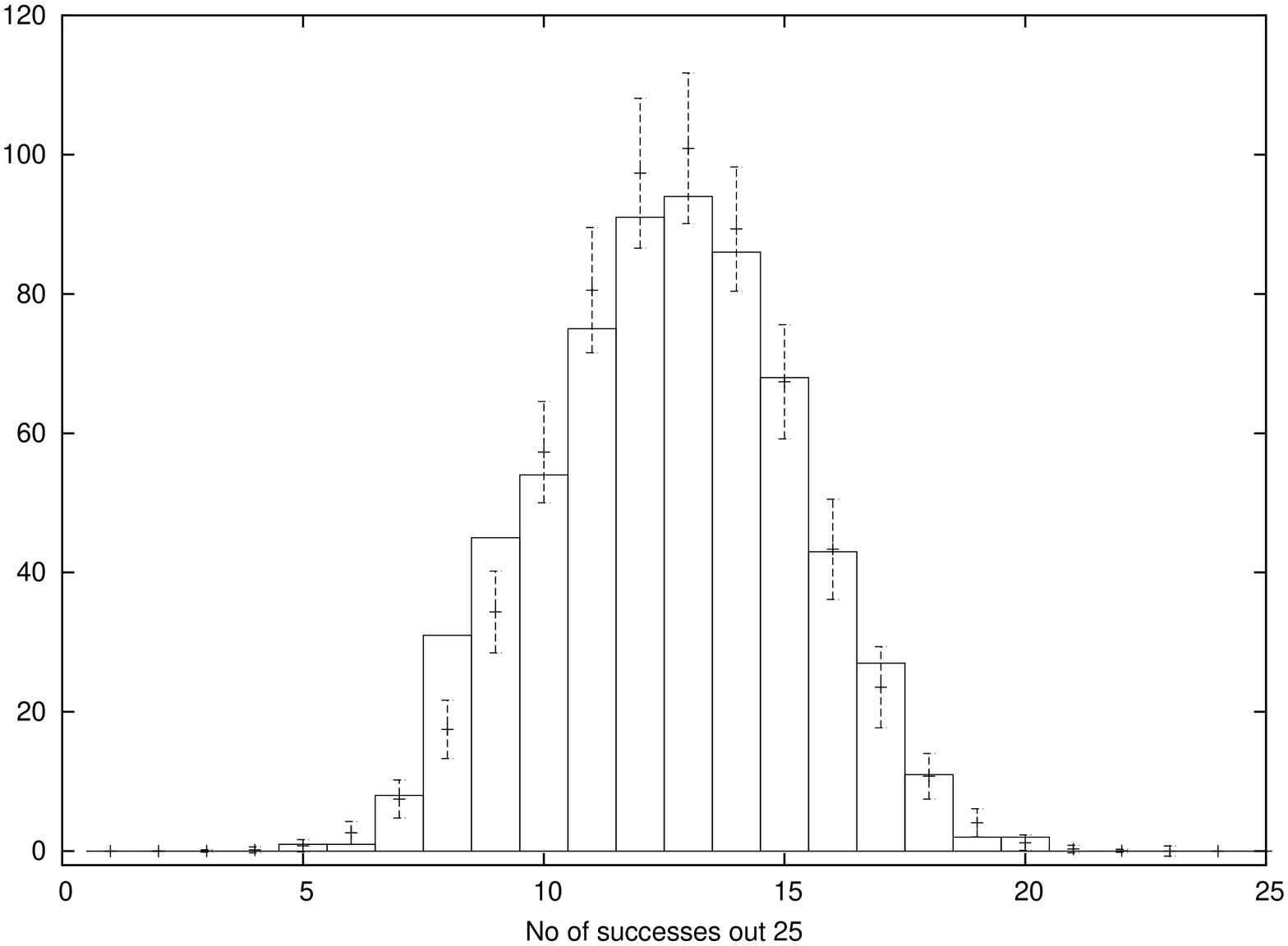,width=8cm,angle=270}
   \psfig{figure=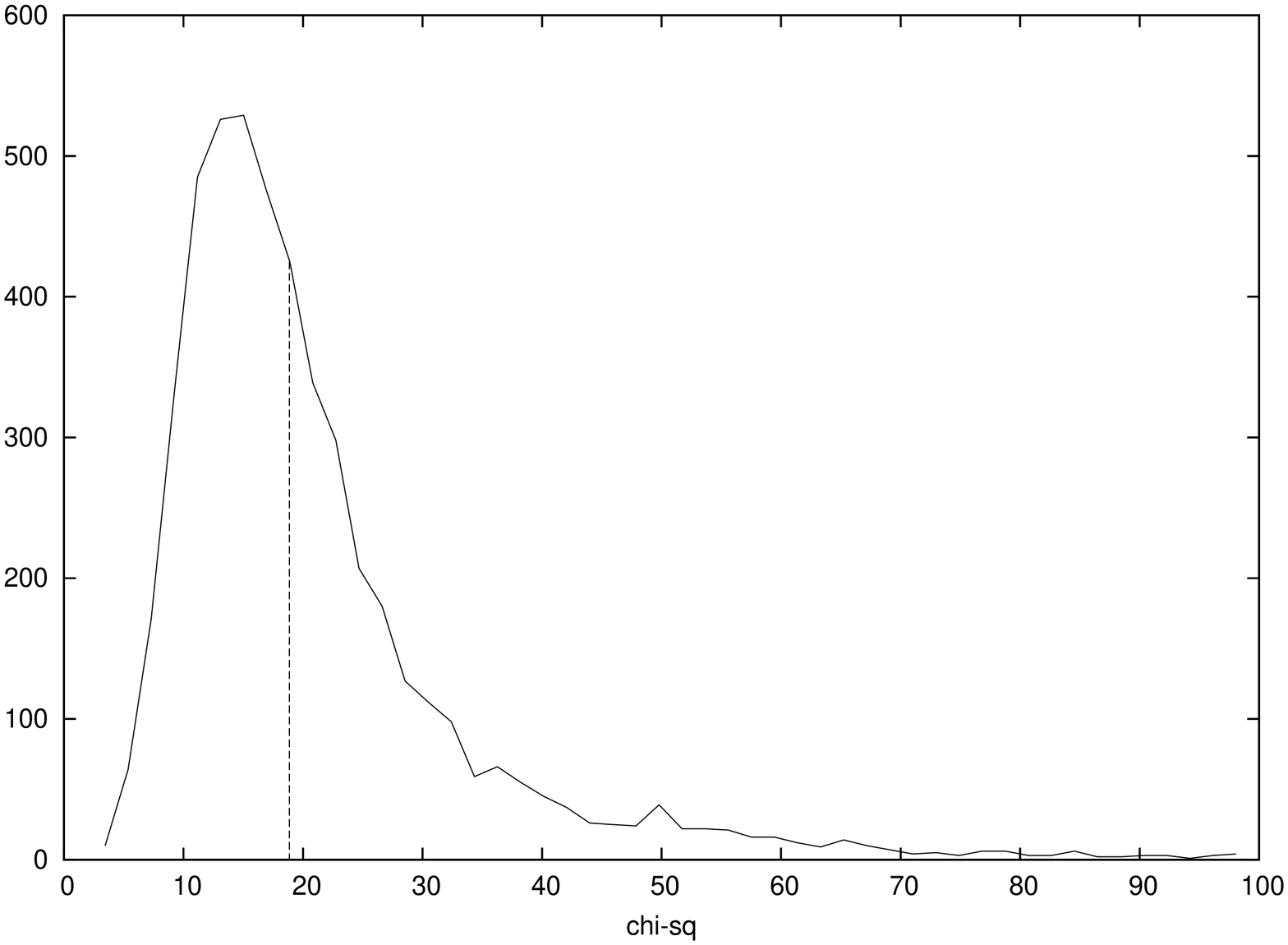,width=8cm,angle=270}\\
 \end{tabular}
\caption{Histograms for low and high redshift samples. The panel on
  the top left shows histogram for objects with low redshift, the
  one at the bottom left is for high redshift objects. The solid line
  represents observed the data and dashed line is for random
  objects. The reduced $\chi^{'2}$ for these two histograms are 0.26
  and 0.29 respectively.} 
\label{fig:hist_z}
\end{figure*}

Hutsem\'ekers et al. have found that the statistical significance of
their alignments are enhanced if they divide their sample into low and
high redshift bins. We do likewise for the JVAS/CLASS results in order
to make as close as possible comparison with results of Hutsem\'ekers
et al.  Out of 4290 objects, redshift information is available for
1273 objects. The redshifts are between 0.03 and 4.72, with median
redshift of 1.24.  Again, the same nearest neighbour test was
performed on the two samples, and on 5000 random realizations keeping
the object positions the same, and assigning the position angles
randomly from a uniform distribution. The histograms with error bars
are shown in Fig. {\ref{fig:hist_z}}, together with the distribution
of $\chi^2$. For the real sky, $\chi^2$ values are 23.8 and 18.8 for
low and high redshift regions, respectively. Values of these or
greater have probabilities of occurrence as almost 1 in 2 and 1 in 4,
in low and high redshift regions, respectively.

\section{An examination of the Hutsem\'ekers et al. regions }
{\label{sec:huts}}

\begin{figure*}
\begin{tabular}{rr}
\psfig{figure=huts_reg1_flat.ps,width=6cm,angle=270}
\psfig{figure=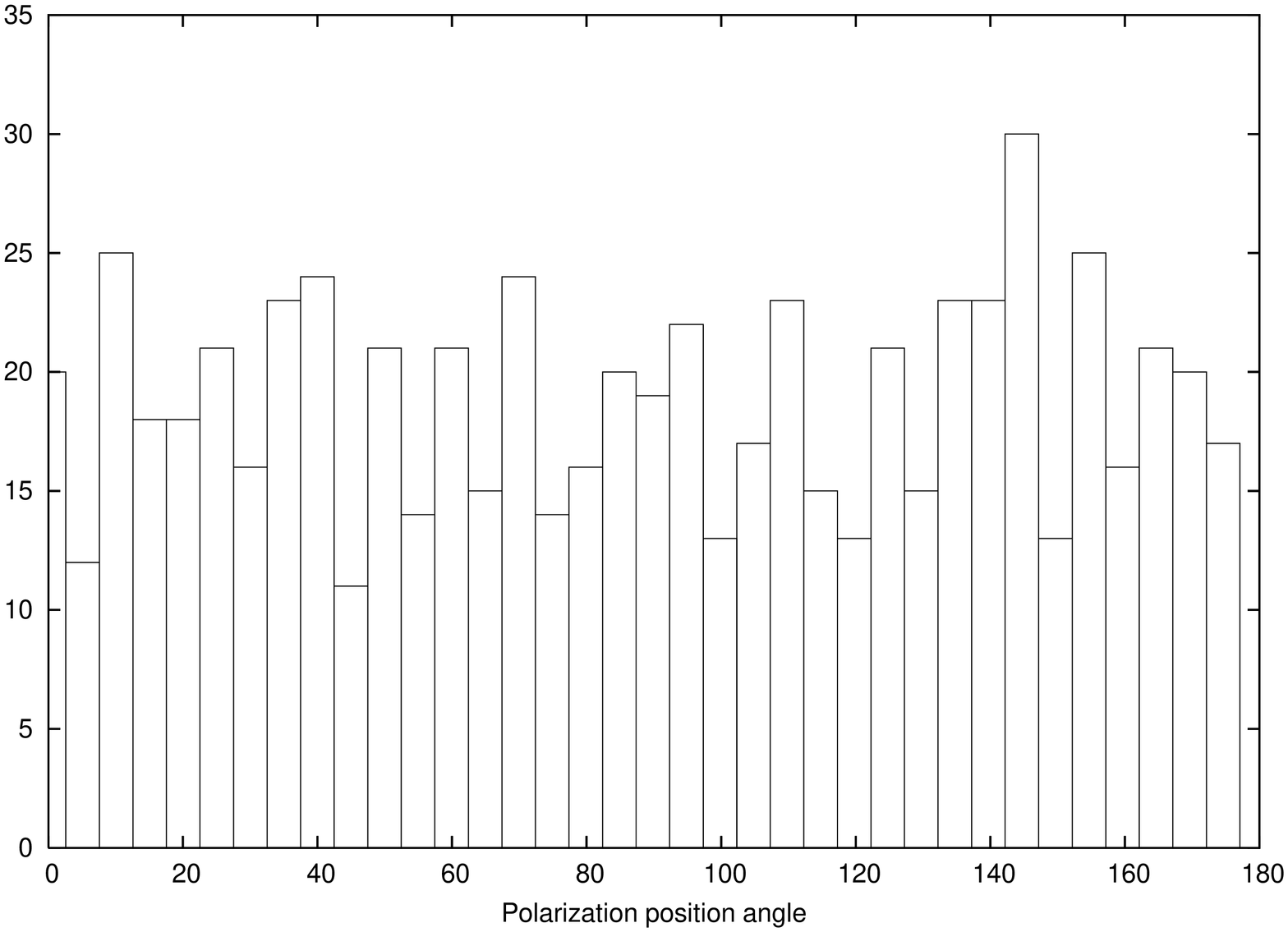,width=8cm,angle=270}\\
\psfig{figure=huts_reg2_flat.ps,width=6cm,angle=270}
\psfig{figure=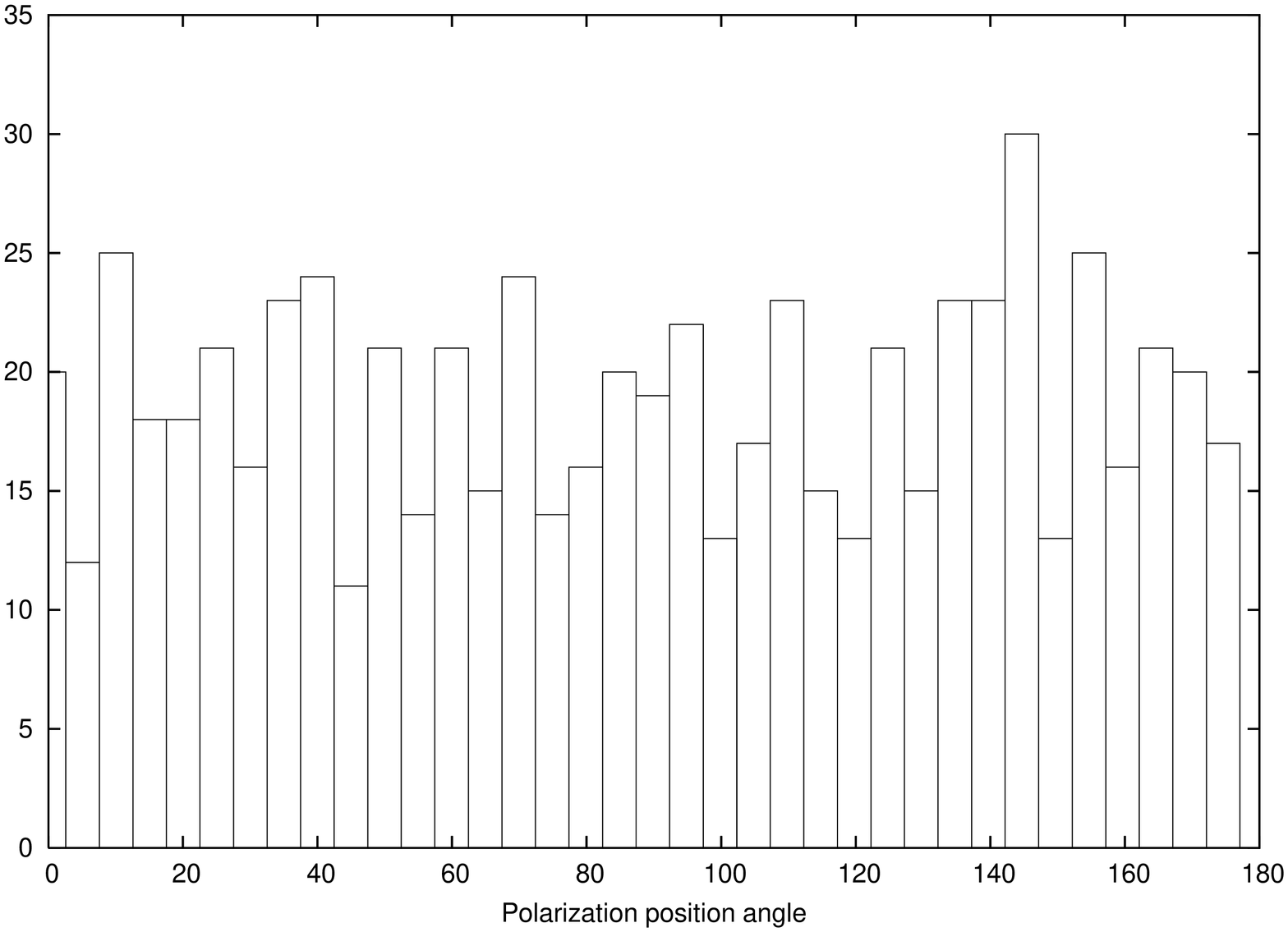,width=8cm,angle=270}\\
\end{tabular}
\caption{The left top panel shows flat-sky projection of the 8.4 GHz
  PAs in Hutsem\'ekers et al.  region A1, and the right hand panel
  shows histogram of the PAs. The bottom panels are the same for
  Hutsem\'ekers et al. region A3.}
\label{fig:HutsRadio}
\end{figure*}

\begin{figure*}
\begin{tabular}{lr}
\psfig{figure=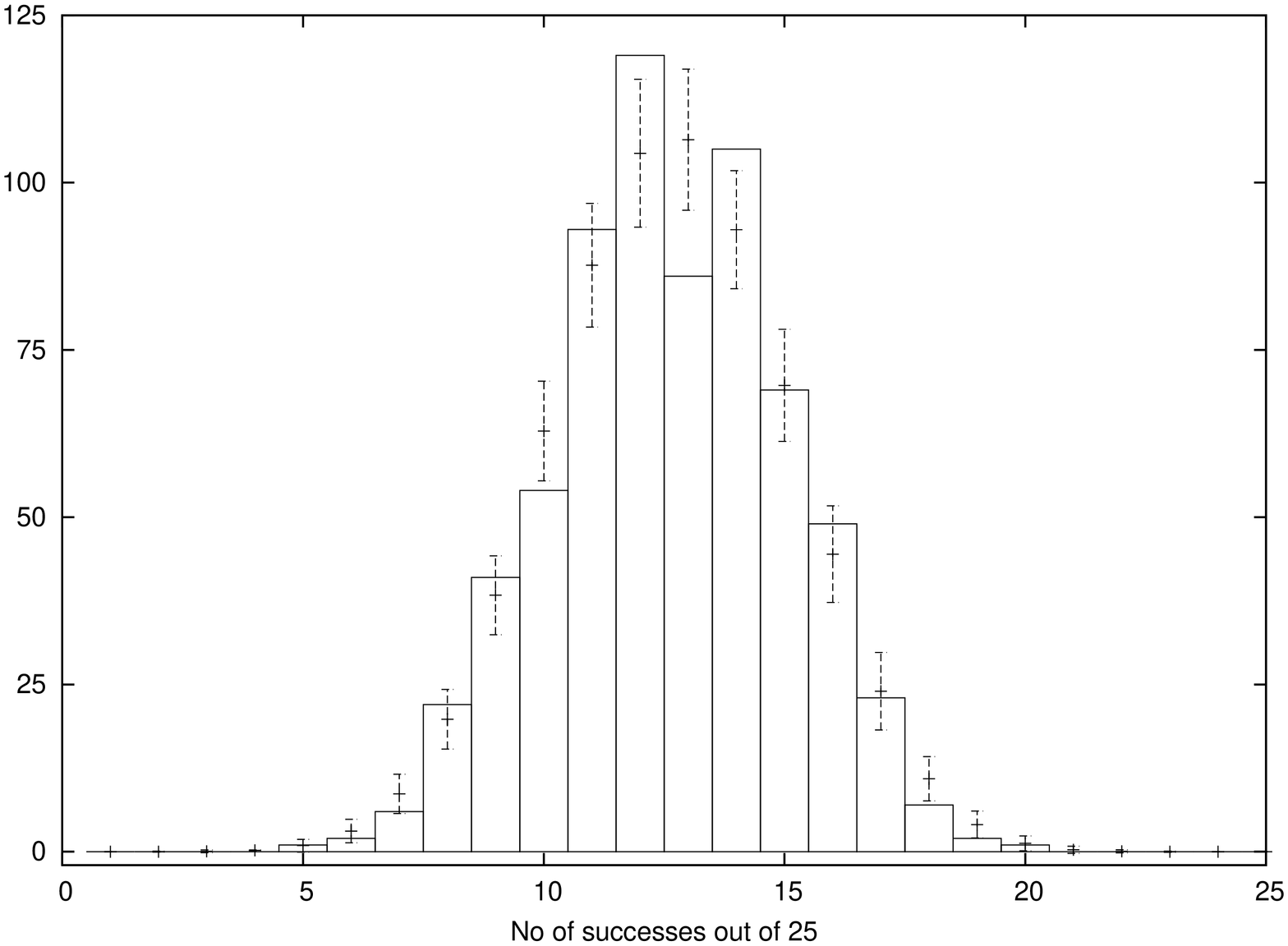,width=8cm,angle=270}
\psfig{figure=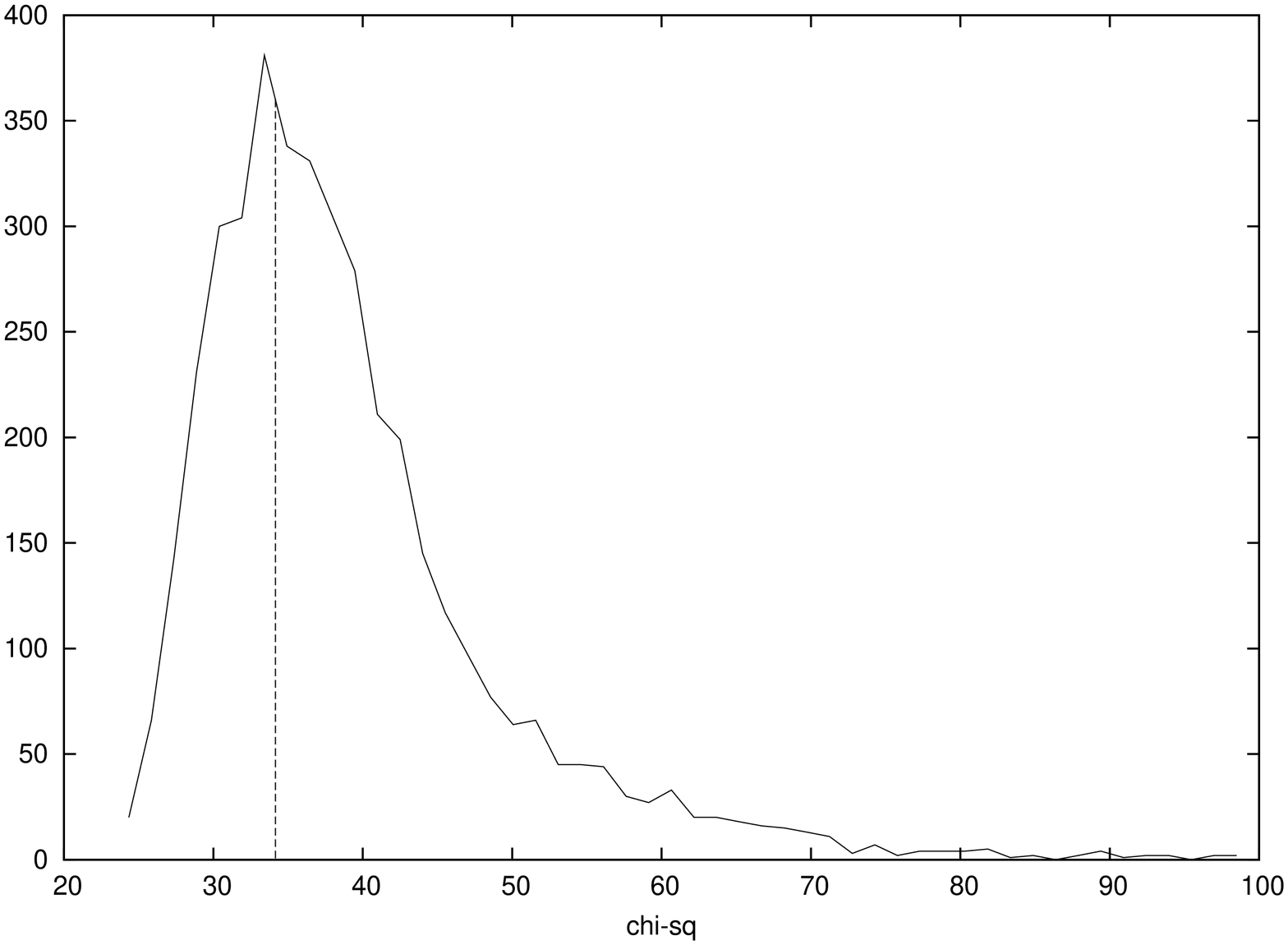,width=8cm,angle=270}\\
\psfig{figure=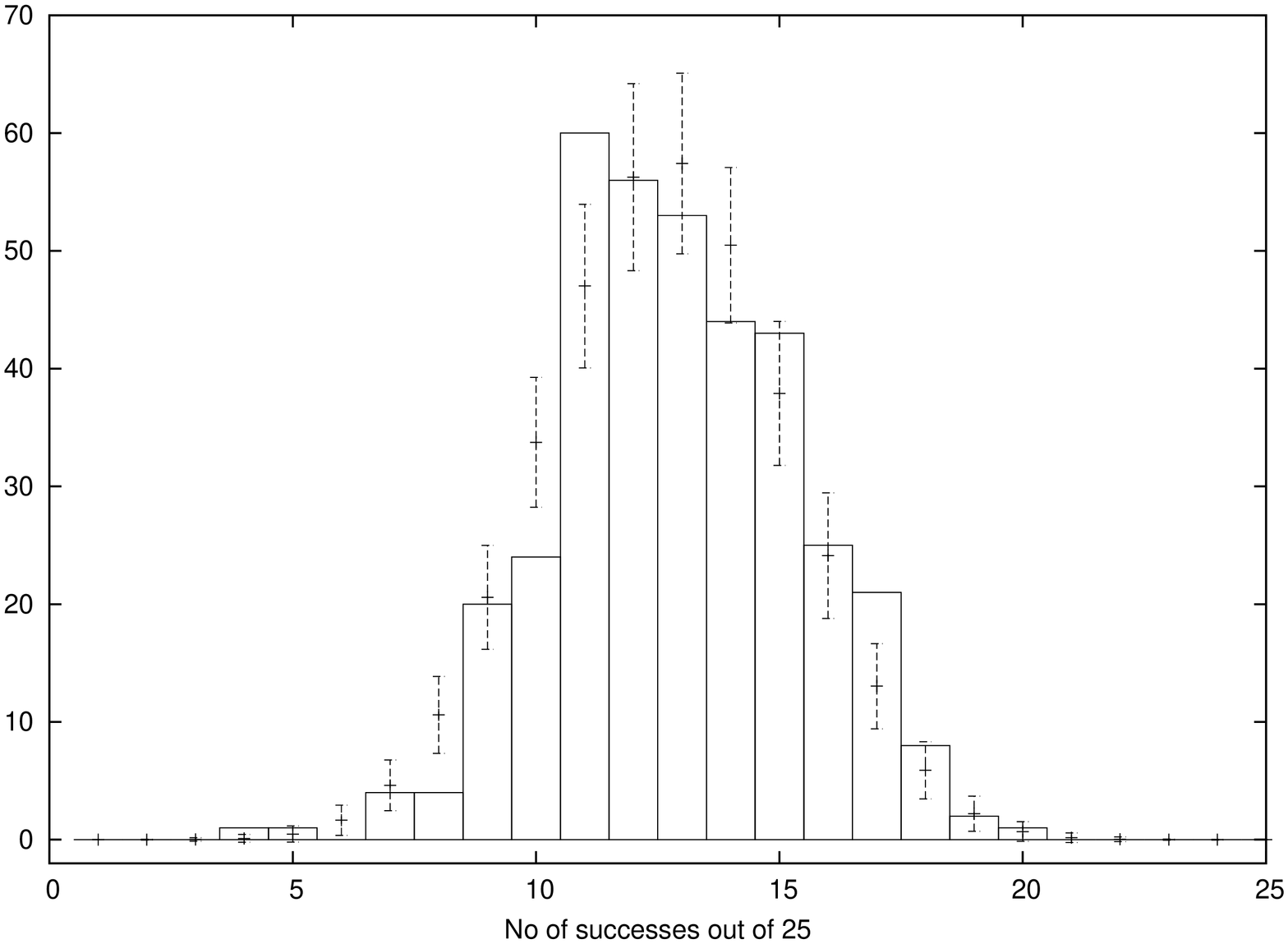,width=8cm,angle=270}
\psfig{figure=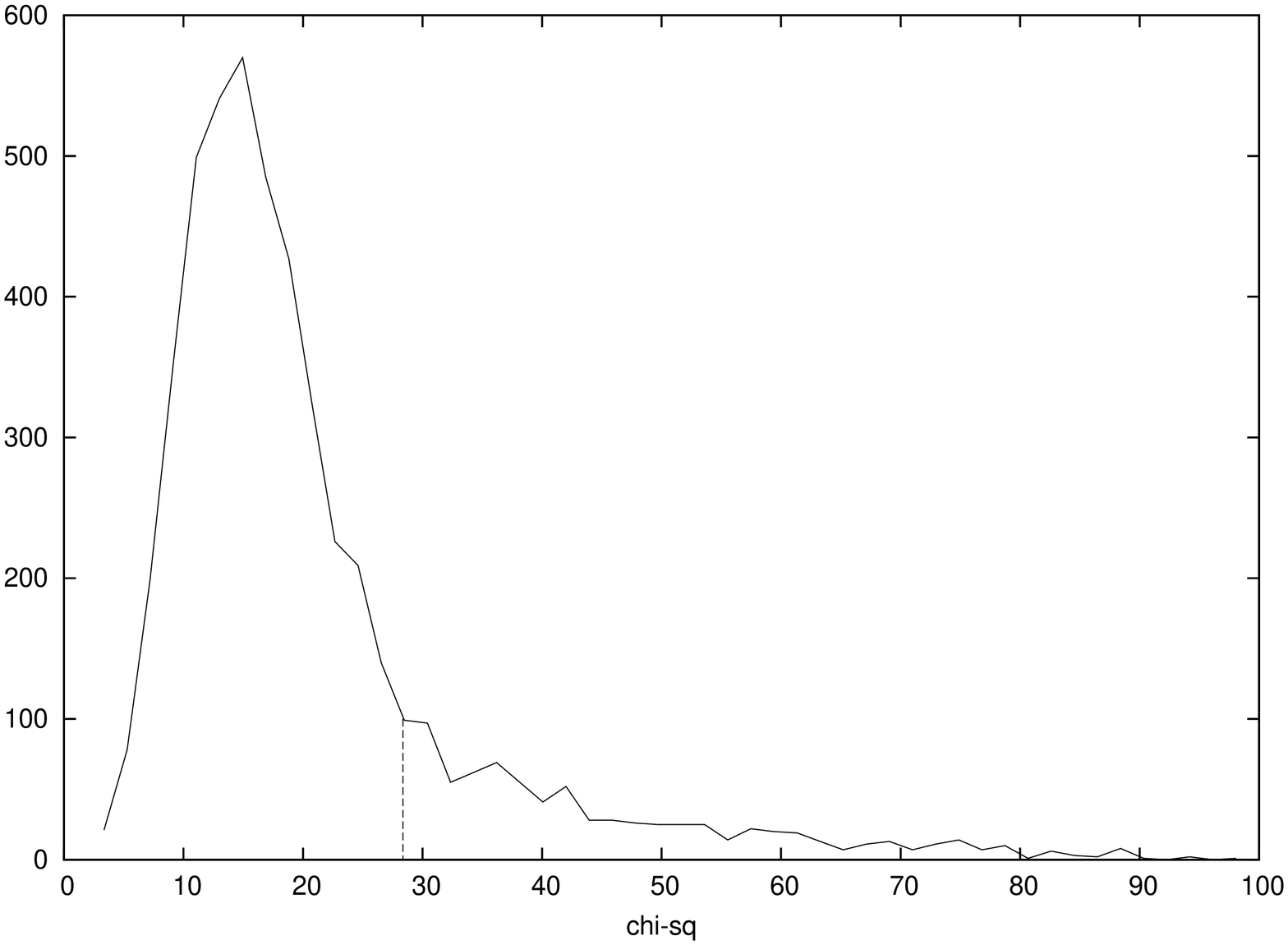,width=8cm,angle=270}\\
\psfig{figure=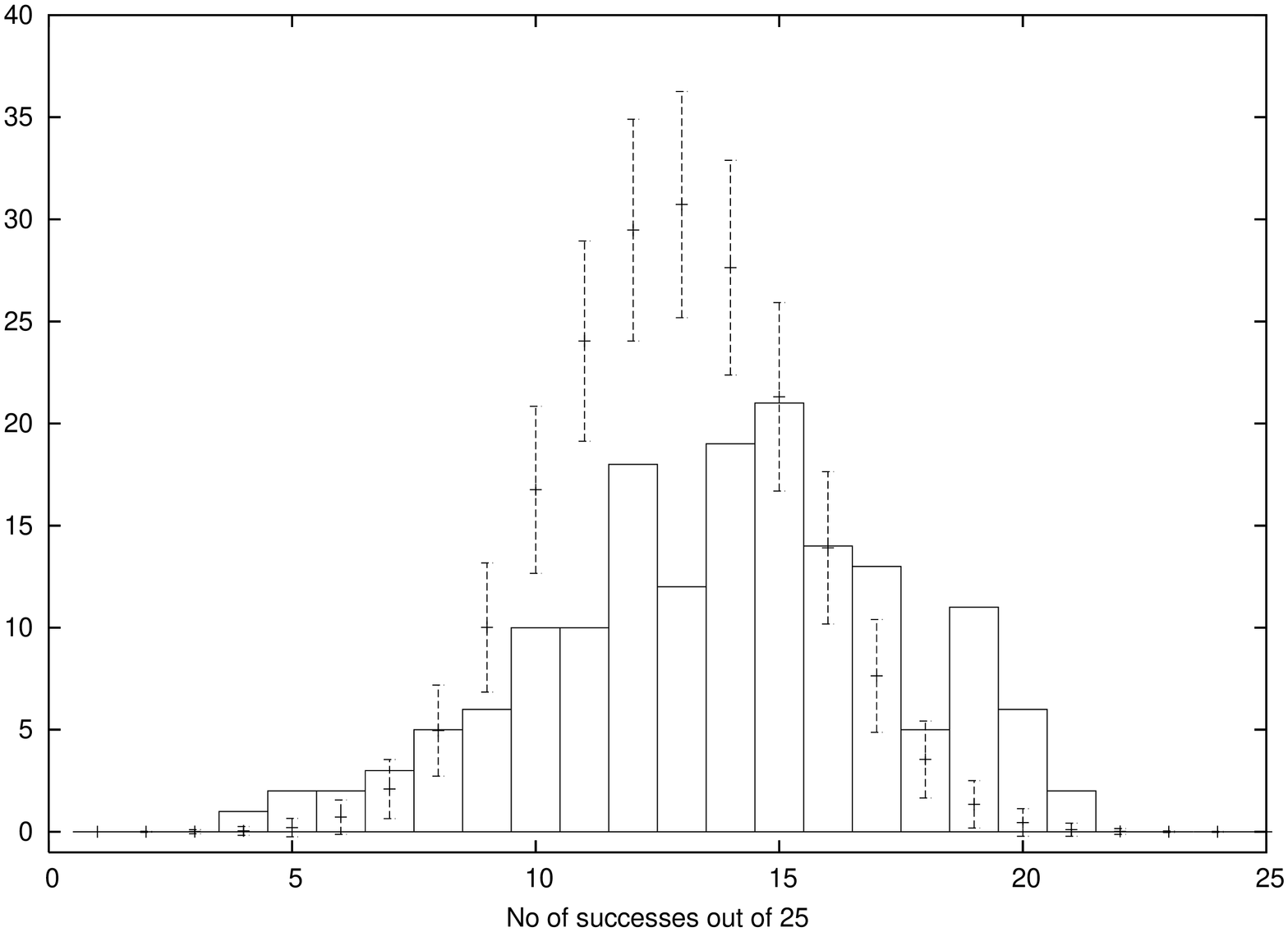,width=8cm,angle=270}
\psfig{figure=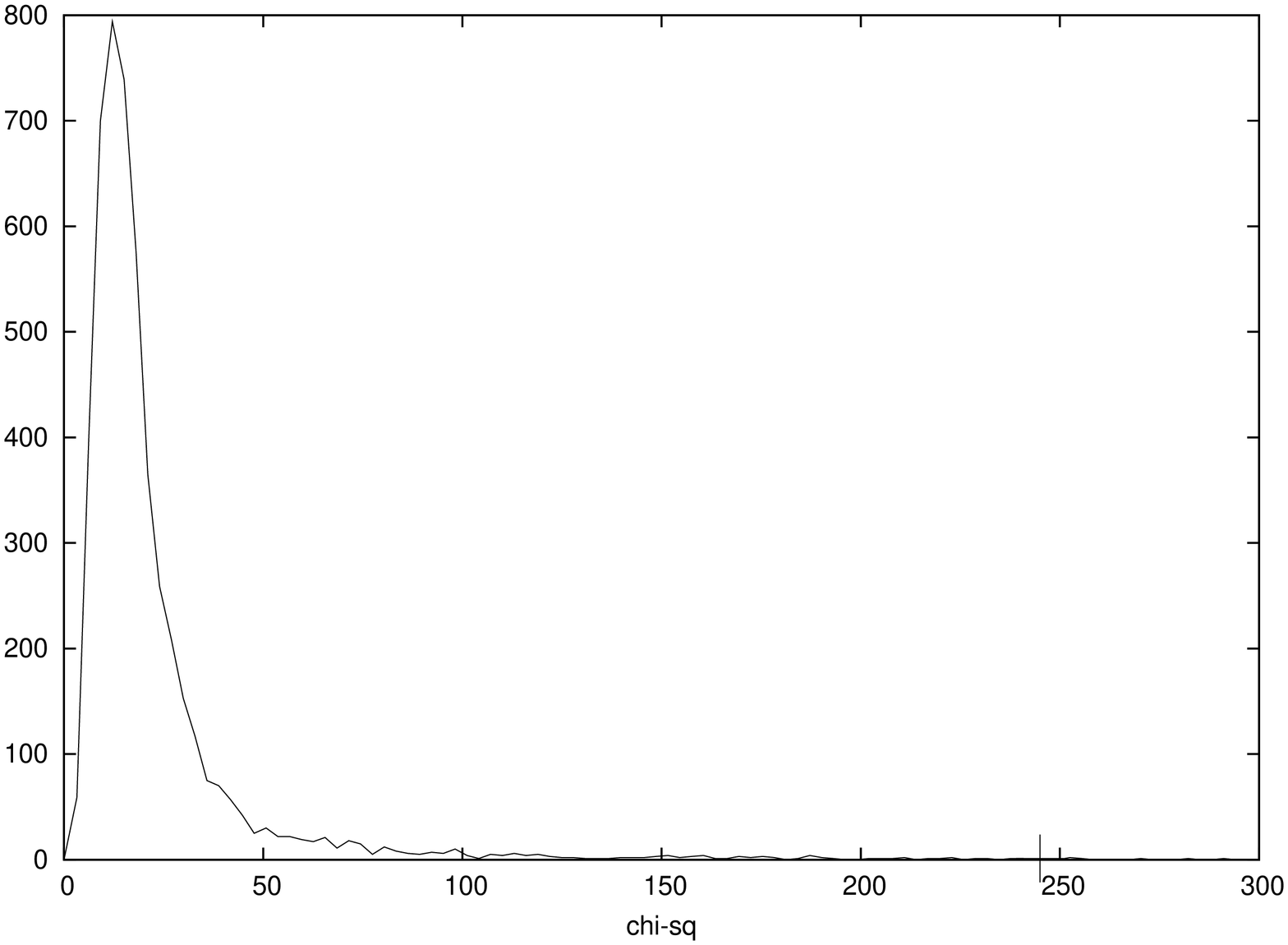,width=8cm,angle=270}\\
\end{tabular}
\caption{The left top panel shows the histogram from nearest neighbour
  test for region A1, with error bars from 5000 random
  realizations. The right top panel shows the $\chi^2$ distribution
  for random realizations with the line showing that for the real sky
  in region A1. The middle panels show the same, for region
  A3. $\chi^2$ for the region A1 is 34.1 and 28.3 for the region
  A3. The probability of finding this $\chi^2$ or more is 52\% for
  region A1 and that for region A3 18\% respectively. The panels at
  the bottom are for region A1 again, but show the results for the
  optical data. For these $\chi^2$ is 245 with probability of 0.7\%
  under the hypthesis that the distribution of position angles is
  uniform.}
\label{fig:hist_huts}
\end{figure*}

Hutsem\'ekers et al. claim that in different regions of sky quasar
optical polarization position vectors are strongly aligned. In Fig
{\ref{fig:HutsRadio}} we plot the distribution of radio polarization
position angles of our sources that are found within the regions
identified by Hutsem\'ekers et al.  They defined regions A1 and A3 as,
168$^\circ$ $\le$ $\alpha$ $\le$ 218$^\circ$ and $\delta$ $\le$
50$^\circ$, and 320$^\circ$ $\le$ $\alpha$ $\le$ 360$^\circ$ and
$\delta$ $\le$ 50$^\circ$, respectively. We do not have any radio data
for $\delta$ $<$ 0$^\circ$ so we do not have exactly the same area
coverage as they do. No obvious alignments are visible on the sky or
in the histograms of position angles also shown in
Fig. {\ref{fig:HutsRadio}}. We have also performed our nearest
neighbour test on the observed data in these regions and again
produced 5000 random realizations as described above. The resulting
histograms are shown in Fig. {\ref{fig:hist_huts}}. The two regions
have $\chi^2$ of 34.1 and 28.3 respectively, with the probability of
occurrence of a $\chi^2$ greater than or equal to these values being
almost 1 in 2 and 1 in 5, respectively. In contrast, our test applied
to the optical data, showed a high significance detection of
non-uniformity. The probabilities of the optical position angles being
uniformly distributed are $\leq$ 0.7\% and 12\%, for regions A1 and
A3, are respectively{\footnote {It should be noted that Hutsemk\'ekers
et al. preferentially targeted quasars in redshift slices where
alignments had been previously suspected and thus the non-uniformity
might be more prominent than in the radio where no redshift targeting
was done.}} (See Fig. {\ref{fig:hist_huts}}, panels at the bottom for
region A1.)

Finally we compare the radio and optical position angles for objects
common to JVAS/CLASS and Hutsem\'ekers et al. The areas do not overlap
completely as Hutsem\'ekers et al., have analyzed the data for
alignments near northern and southern galactic poles, whereas
JVAS/CLASS surveys covered all RA ranges but only declination $\ge$
0$^{\circ}$. We find 52 of their sources for which we have reliable
polarization measurements at 8.4 GHz. (See appendix for the data.)
Figure {\ref{fig:hist_comm}} shows the distribution of PA differences in
optical and radio. There is no obvious correlation between the optical
and radio polarization position angles. This is perhaps not surprising
since correlations between radio and optical polarization position
angles have been searched for in the past but none conclusively found
(Lister \& Smith (2000)). In summary, we find no evidence to
suggest that there is anything special about the radio polarization
properties of the objects found in the Hutsem\'ekers et al. regions.

\begin{figure*}
\psfig{figure=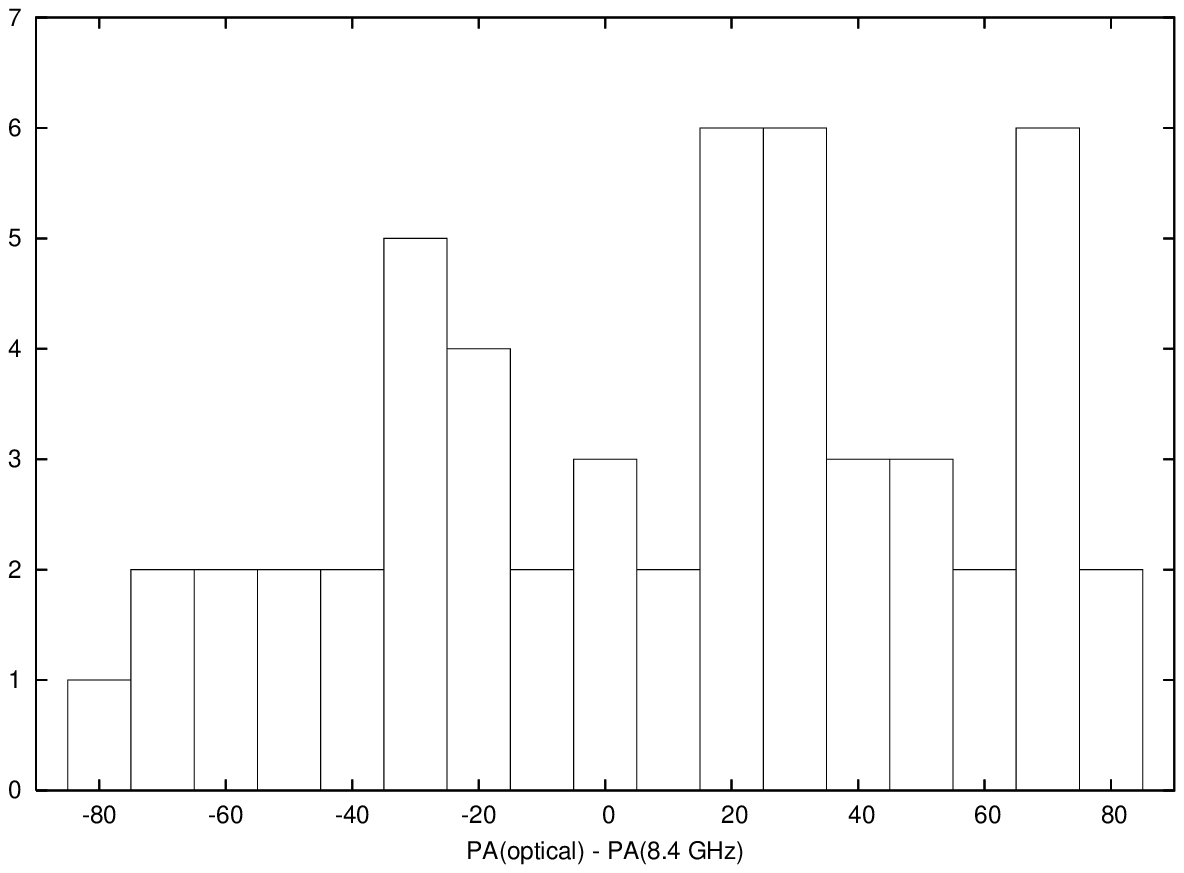,width=8cm,angle=0}
\caption{The PA difference distribution for 52 common sources in
  JVAS/CLASS survey and Hutsem\'ekers et al (2005) sample of quasars.}
\label{fig:hist_comm}
\end{figure*}

\section{Comparison between polarization position angles and the axes
  of parsec-scale jets} 

Birch (1982) claimed that there were systematic offsets between radio
polarization position angles and the direction of elongation of radio
sources and that these were systematically correlated over the large
regions.  We have undertaken a similar investigation using our
polarization position angle measurements and jet axes derived from
VLBI maps. We use information on sub-kiloparsec scales rather than
much larger scale information used by Birch because the majority of
JVAS/CLASS objects, as a result of the spectral index selection used to
define the sample, are compact with little observational information
available on their structures on the scales of 10s to 100s of
kiloparsecs. However, there is plenty of parsec-scale
information. Using data from the literature (Beasley et al., 2002;
Fomalont et al., 2003; Henstock et al., 1995; Kovalev et al., 2005;
Petrov et al., 2005; Petrov et al., 2005; Kovalev et al., 2007;
Polatidis et al., 1995; Taylor et al., 1994; Thakkar et al., 1995; Xu
et al., 1995; Zensus et al., 2002) and that available in web-based
archives we have estimated jet position angles for 1565 sources. There
are 842 of these for which we also have polarization position
angles. (See online data.) The jet position angle is defined to be in
the range $-180^{\circ}$ to $180^{\circ}$ measured from North through
East. The position angles were measured by eye and we estimate the
likely uncertainty of these measurements to be $\pm10^{\circ}$.

\begin{figure*}{\label{fig:JetAngDist}}
\begin{tabular}{lr}
\psfig{figure=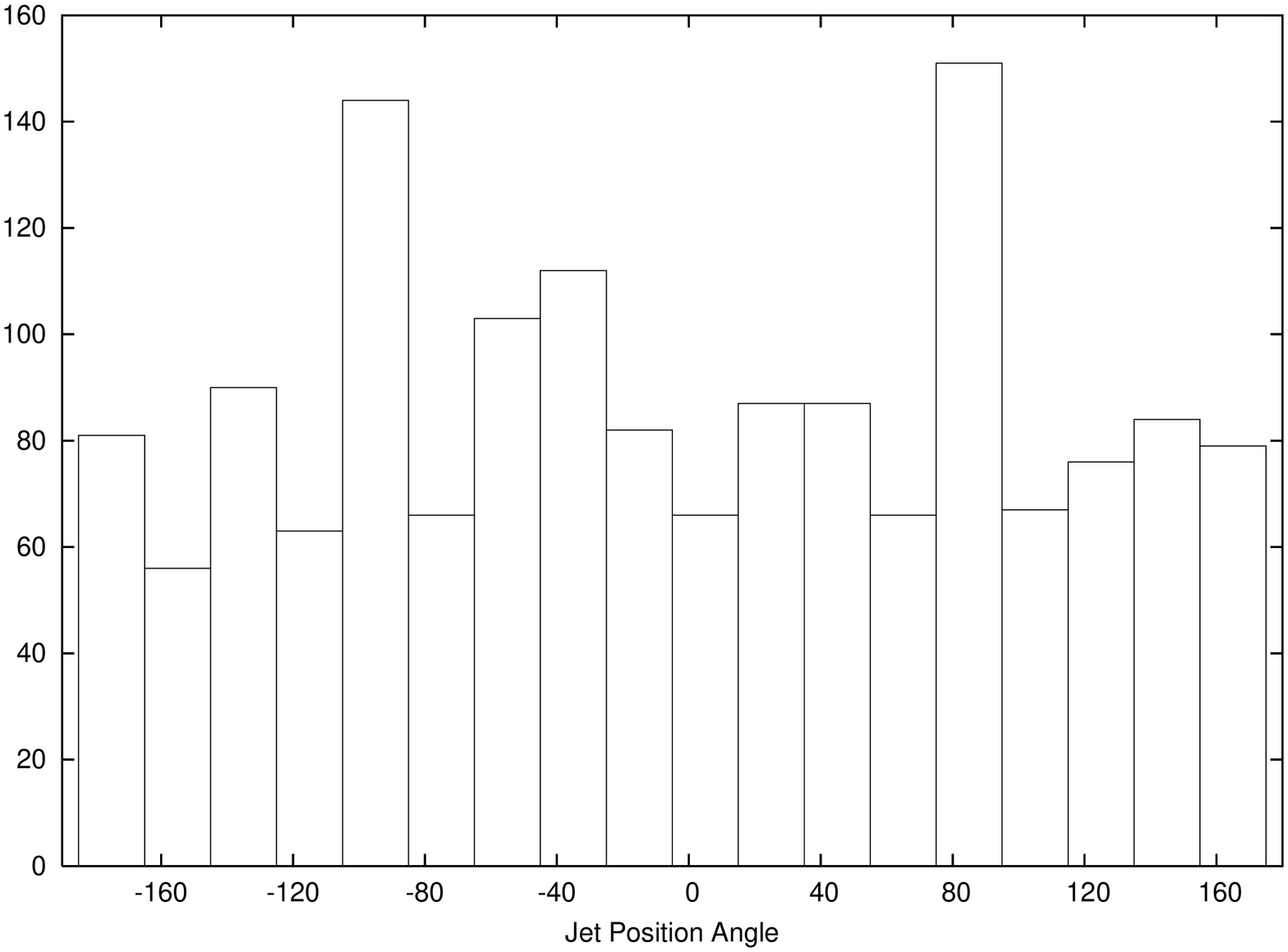,width=8cm,angle=270}
\psfig{figure=jetang_aitoff.ps,width=8cm,angle=270}
\end{tabular}
\caption{a) The left hand panel shows jet-angle distribution of 1565
  sources. Two peaks at $\pm$ 90$^\circ$ stand out clearly. b) On the
  right hand is the Aitoff projection of distribution of jet PA on the
  sky.} 
\end{figure*}

The first thing to note is that the distribution of jet position
angles is not uniform (see Fig. 9a). The two peaks at $\pm$ 90$^\circ$ 
arise because typical VLBI arrays have more resolution in the
East-West direction than they do in the North-South direction. There
is, therefore, a higher probability of being able to detect a
measurable jet in the East-West direction than in the North-South.
We note, however, that the distributions of jet angles should not
depend on the areas of sky in which they are measured. The non-uniform
overall distribution should not severely affect local statistics such
as those employed in the earlier analysis of polarization position
angles.

In order to investigate local correlations, we have done two things:
we have examined the distribution of position angles to see if there
are any systematic deviations from uniformity in different parts of
the sky and we have looked at the polarization position angle/VLBI jet
angle difference for any similar effects. 

\subsection{The distribution of jet position angles} 

We have performed the histogram and random walk tests on the data
using $n_{\rm bins}=4$ and the celestial coordinate system. The
results are presented in Table {\ref{tab:jets}. It is clear that for
$\theta_{\rm pix}\le 29.3^{\circ}$ the results are compatible with
uniformity and the case of $\theta_{\rm pix}=58.6^{\circ}$ is
dominated by small number statistics.

\begin{table*}
\begin{center}
\begin{tabular}{|cc|cc|cc|}
\hline
 $\theta_{\rm pix}/{\rm deg}$ & $n_{\rm obj}$ & $N_{\rm hist}(95\%)$ & $P_{\rm hist}(95\%)$ & $N_{\rm rand}(95\%)$ & $P_{\rm rand}(95\%)$ \\
\hline
58.5 & 12 & 1 & 8.3 & 2 & 16.0 \\
29.3 & 44 & 0 & 0.0 & 1 & 2.3 \\
14.7 & 156 & 4 & 2.6 & 6 & 3.9 \\
7.3 & 549 & 2 & 0.36 & 17 & 3.1 \\
\hline
\end{tabular}
\end{center}
\caption{Results of using the histogram and random walk tests on the
  jet position angle data. The first two results are the number and
  percentage of pixels which fail the histogram test at 95\% and the
  second two are for the random walk test.} 
\label{tab:jets}
\end{table*}

We also have performed the nearest neighbour test on the jet angle
data to look for regions of sky where there might be clusters of
sources with aligned position angles. No obvious regions were found.
From these three tests, therefore, we conclude that
the jet angles show no tendency to align in different regions of sky.

\subsection{Correlations between jet and polarization position angles}

The histogram of the difference between jet PA and polarization PA is
shown in Fig. 10. It is clear that there is a peak near 90$^\circ$
with a less significant peak around 0$^\circ$. These 90$^\circ$ and
0$^\circ$ peaks are consistent with previous results (e.g. Gabuzda et
al., 1994; Helmboldt, et al., 2007; Lister and Smith 2000; Pollack
et al., 2002). Such a correlation is also expected on astrophysical
grounds since the magnetic field vectors of the emitted radiation are
known to align along the local jet direction or, less frequently,
perpendicular to it (e.g. Gabuzda et al., 1992). The lack of
uniformity in the polarization/jet position angle distribution, which
survives the effects of resolution and selection involved in the
construction of our sample, means that one needs to be careful when
looking for signatures of systematic alignments. However, an obvious
test is to look at the sign of the position angle difference for any
evidence of an effect similar to that found by Birch (1982). We have
done this using the nearest neighbour test where we define a 'success'
as a neighbour having the same sign as the source in question and a
failure, one having the opposite sign. The results of the test are
displayed in histogram form in Fig {\ref{fig:jeta}}. There is no
evidence for regions having large numbers of either negative or
positive position angle differences. 
\begin{figure*}{\label{hist_ja_pa}}
\psfig{figure=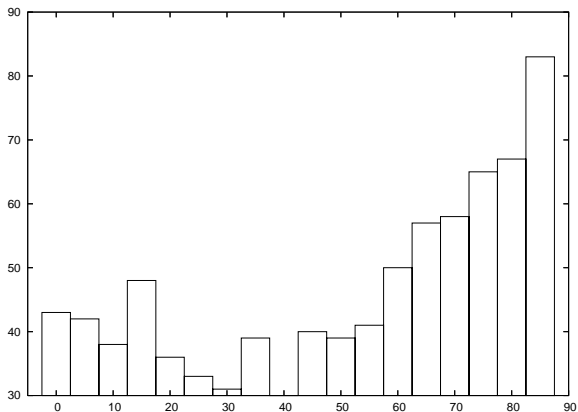,width=8cm,angle=270}
\caption{The histogram of difference between VLBI jet PA and
  polarization PA.}
\end{figure*}

\begin{figure*}
\begin{tabular}{lr}
\psfig{figure=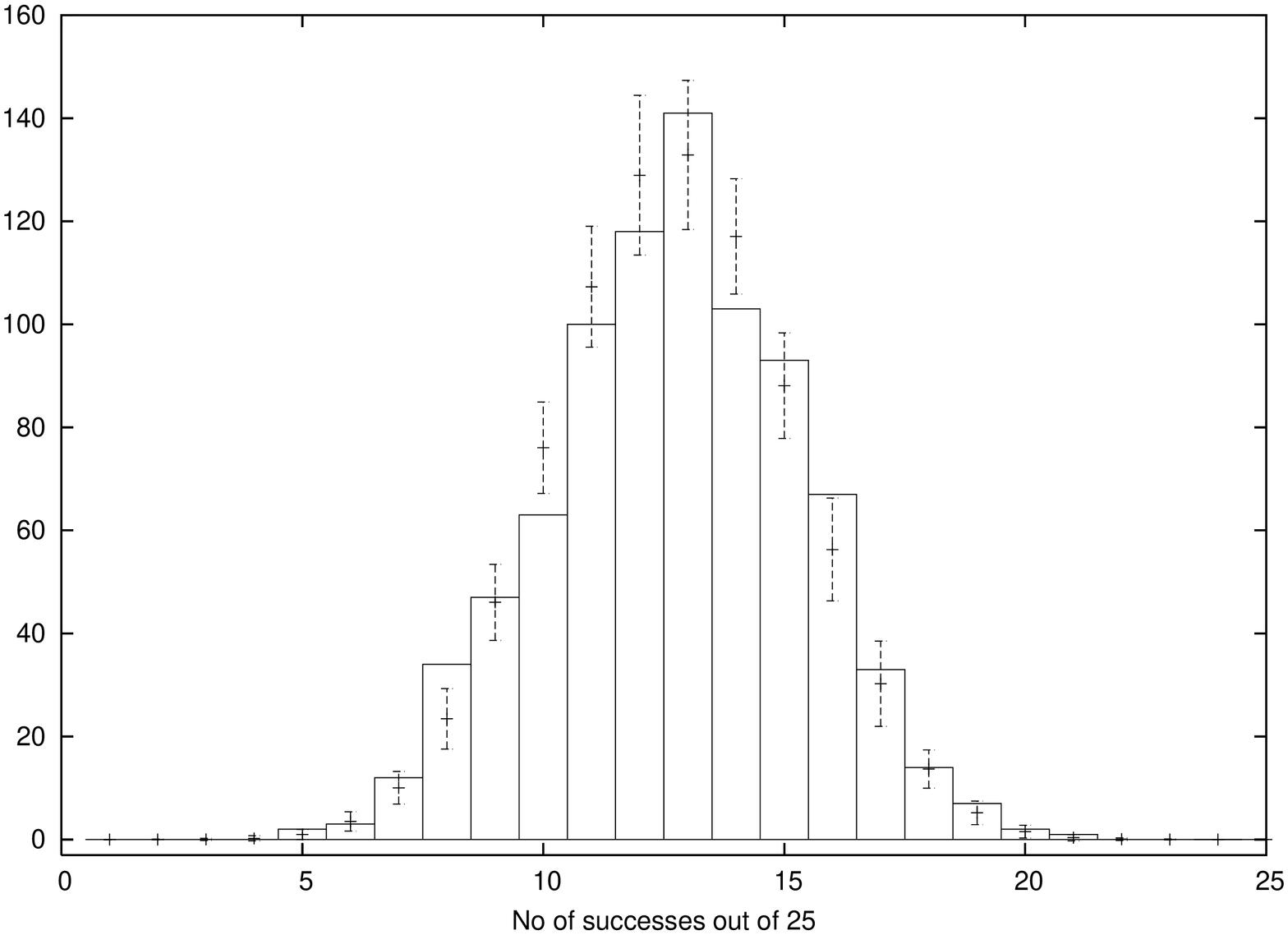,width=8cm,angle=270}
\psfig{figure=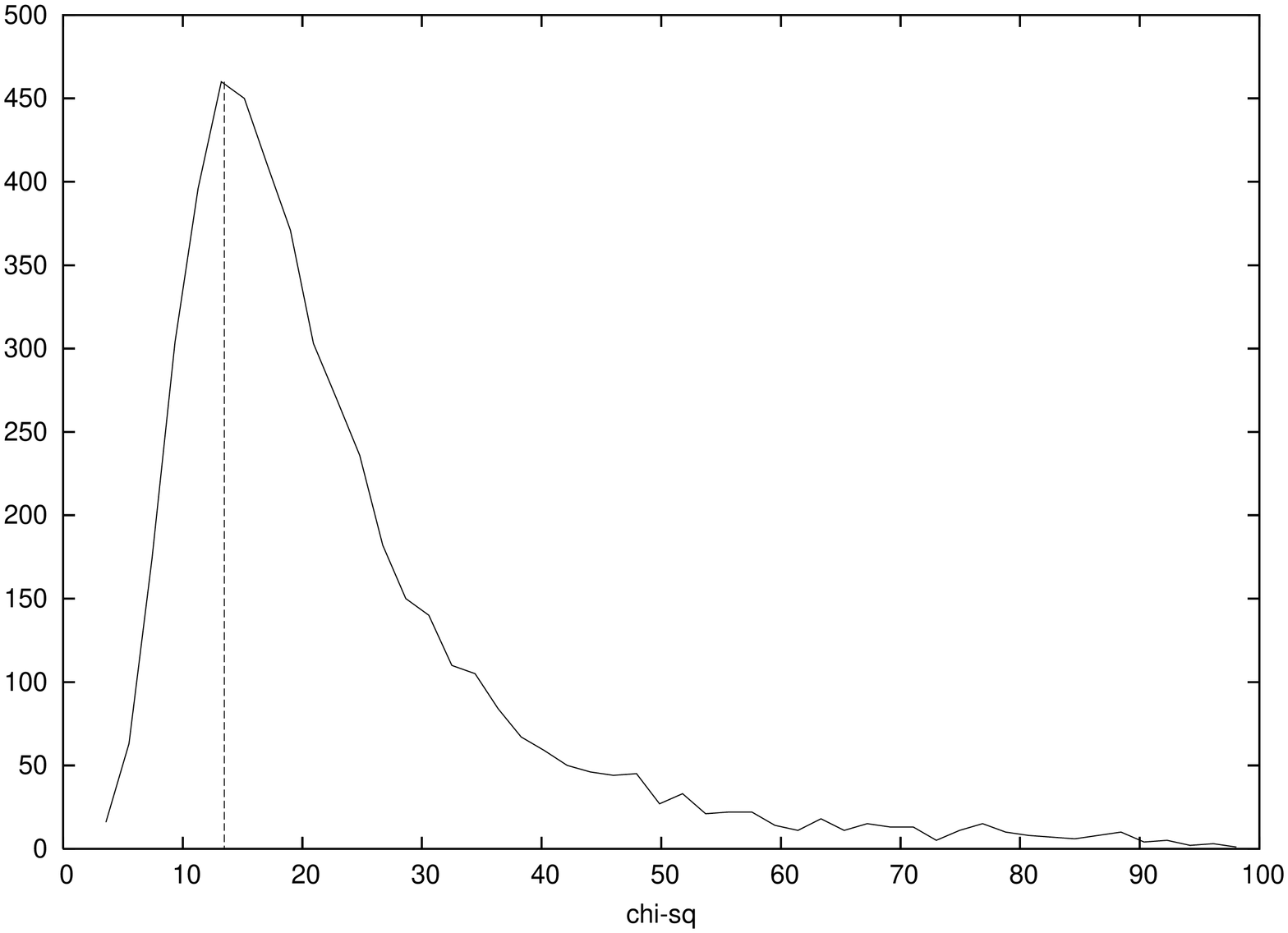,width=8cm,angle=270}
\end{tabular}
\caption{The left hand panel shows the results of the nearest
   neighbour test on differences between jet angles and polarization
   position angles. The right hand panel shows distribution of
   $\chi^2$ with the real data having $\chi^2$ = 13.5, with
   probability of 69\% of occurring under the hypothesis of a uniform distribution.} 
\label{fig:jeta}
\end{figure*}

\section{Discussion}{\label{discussion}}

None of the statistical tests we have performed (see Section
{\ref{stat}}) indicate any areas in which there are large-scale
alignments of PAs measured at 8.4GHz, in contrast to what has been
found in the optical wavelengths by Hutsem\'ekers et al. Could this be
due to some error in the radio results arising from instrumental bias
in PA measurement, or a data analysis error? We think this is highly
unlikely because the data have been carefully re-analyzed paying
particular attention to the polarization calibration and the
extraction of the polarization parameters. External comparisons with
other data, including our own Westerbork observations, convince us
that CLASS polarization angles are reliable (See Section
{\ref{WSRT-text}}.) 

We can think of two generic reasons why there might be alignments
found in optical polarization position angles and not the radio ones: 

\begin{itemize}

\item The physical mechanism that aligns the optical polarizations may
  be frequency dependent and not apply to the radio. This could either
  apply to the production of the radiation or any polarization imposed
  during propagation. 

\item The optical results could be misleading because of either
  statistical fluke or the polarizations are not primarily
  extragalactic in origin. 

\end{itemize}

We first consider physical mechanisms. We have checked the optical
sample of 355 quasars using V\'eron \& V\'eron (2007) and the NASA
Extragalactic Database (NED) and find that $\sim$72\% are radio loud
and most have flat radio spectra. Having a flat radio spectrum
suggests that they are blazars and in blazars the optical and radio
emission is predominantly
synchrotron in origin. Thus for the majority of the radio and optical
objects the radiation mechanism is the same and this argues against
something intrinsic producing alignments detectable at
optical wavelengths and not at radio wavelengths{\footnote {Despite the
common radiation mechanism we and other authors do not find any clear
correlation between the radio and optical position angles. We have
52 objects in common with Hutsem\'ekers et al (see Section {\ref{sec:huts}})}}.

For radio-quiet quasars, the source of intrinsic optical polarization
is likely to be scattering (Stockman, Moore \& Angel (1984), Berriman
et al (1990)). Perhaps significantly, the degree of alignment in
radio-loud quasars and radio-quiet quasars cannot be distinguished
statistically, though it should be noted that the radio-quiet sample
is small. Thus, assuming the optical result is of cosmological
significance, the more plausible explanation would be a propagation
effect that polarizes optical emission and not the radio, or perhaps
one that destroys the radio alignments. An obvious mechanism to
destroy radio alignments is Faraday rotation.  However, rotation
measures would have to be of the order of thousands of radm$^{-2}$
and this possibility can also be ruled out because: 

\begin{itemize}

\item The rotation measures of extragalactic sources in general are
  normally quite low, often less than a few tens of radm$^{-2}$
  (e.g. Rudnick \& Jones (1983), but see also Zavala \& Taylor (2004)
  who measure very high rotation measure but on parsec scales). For
  JVAS/CLASS the random subset of objects for which we have 
  measurements done at 8.4 GHz, 5 GHz (WSRT), and 1.4 GHz (NVSS) the
  rotation measures are around tens of radm$^{-2}$ or less, rather
  than the order of thousands. (Joshi et al, in prep).

\item There is a correlation between the position angles of radio jets 
   and the corresponding magnetic field directions inferred from radio 
   polarization measurements (Wills et al., (1992); Visvanathan \&
   Wills (1998); Fig. 10 of this paper). Such a correlation would not
   be seen if the radio polarizations were significantly randomized by
   Faraday rotation.

\end{itemize}

We conclude that intrinsic mechanisms or Faraday rotation cannot
account for the lack of radio alignments and therefore the most viable
explanations are that either there is some, as yet uncertain,
propagation effect that works on the optical photons and not the
radio, or that the optical results have may not been interpreted
correctly. This might be a result of the relatively small numbers
alone or from the combination of small numbers and any  residual biases in
the data, perhaps caused by the effects of Galactic interstellar
polarization. One suggestion discussed in Hutsem\'ekers et al. (2005)
is that radiation propagating over cosmological distances can be
polarized by light weakly interacting pseudoscalar (or scalar)
particles (Das et al, 2005) and that this propagation effect can be
frequency-dependent.  However, given the profound implications of a
non-Galactic astrophysical origin for the observed optical result, and
the absence of evidence for the equivalent effect seen in the radio,
we remain unconvinced by the conclusions of Hutsem\'ekers et al.

Finally, since there are models which predict that polarization
PA could suffer small ($\sim1^{\circ}$) systematic rotations on large
angular scales (e.g. Skrotskii rotation produced by vector
perturbations of a flat Friedmann-Robertson-Walker background
described in Morales \& S\'aes, (2007)), we briefly address the 
question of whether or not in the future it might be possible to
detect such small deviations from random alignments using radio
polarization data. We suggest that the most promising approach would
be to use the difference in PA between that of the
polarization and that of parsec-scale jets measured with VLBI. The two
are observed to be correlated with a dispersion of
$\sim$30$^{\circ}${\footnote {The smaller dispersion relative to the
raw polarization position angles means that smaller samples would be
required to detect an effect.}}  and only the polarization angles
should be affected by the Skrotskii rotation.  Approximately a
thousand objects would be required to constrain the mean position
angle at the degree level (30$^{\circ}/\sqrt{1000}$).  To detect an
effect, several regions containing the order of 1000 objects would be
required. Thus, assuming no systematic errors, in samples of
$\sim10^4$ objects one would begin to be sensitive to alignments at
the predicted levels. Already the number of objects with both
polarization position angle measurements and jet measurements are
significantly in excess of a thousand, so such tests of this type are
beginning to be feasible and will be very powerful in the SKA era.

\section*{Acknowledgements}

We thank numerous colleagues for advice, particularly Paddy Leahy and
Simon Garrington. The National Radio Astronomy Observatory is a
facility of the National Science Foundation operated under cooperative
agreement by Associated Universities, Inc. The Westerbork Synthesis
Radio Telescope is operated by the ASTRON (Netherlands Foundation for
Research in Astronomy) with support from the Netherlands Foundation
for Scientific Research NWO. This research has made use
of the NASA/IPAC Extragalactic Database (NED) which is operated by the
Jet Propulsion Laboratory, California Institute of Technology, under
contract with the National Aeronautics and Space Administration.  This
work was supported in part by the European Community's Sixth Framework
Marie Curie Research Training Network Programme, Contract No.
MRTN-CT-2004-505183 "ANGLES". S.J. thanks the University of Manchester
School of Physics and Astronomy for support.

\section*{References}

\noindent Beasley, A.J., Gordon, D., Peck, A.B., Petrov, L.,
MacMillan, D.S., Fomalont, E.B., Ma, C., 2002, ApJS, 141,
13 

\noindent Berriman G., Schmidt G.D., West S.C., Stockman H.S., 1990,
ApJS, 74, 869. 

\noindent Birch P., 1982. Nat 298, 451.

\noindent Browne I.W.A., Wilkinson P.N., Patnaik A.R. Wrobel J.M.,
1998, MNRAS, 293, 257.

\noindent Browne I.W.A., Wilkinson P.N., Jackson N.J.F., Myers S.T., 
Fassnacht C.D., Koopmans L.V.E., Marlow D.R., Norbury M., Rusin D., 
Sykes C.M.,  2003. MNRAS 341, 13.

\noindent Carroll S. M., and Field G. B., 1997, Phys. Rev. Lett., 79, 2349

\noindent Das, S., Jain, P., Ralston, J. P., Saha, R., 2005, JCAP, 6, 2

\noindent Eisenstein D. J., and Bunn E. F., 1997, Phys. Rev. Lett.,
79, 1957

\noindent Falco E., Kochanek C. S., Munoz J. A., 1998, ApJ, 494, 47. 

\noindent Fomalont, E., Petrov, L., McMillan, D.S., Gordon, D., Ma,
C., 2003, AJ, 126, 2562

\noindent Gabuzda D.C., Rastorgueva E.A., Smith P.S., O'Sullivan S.P.,
2005, MNRAS, 369, 1596 

\noindent Gabuzda D.C., Mullan, C. M., Cawthorne, T. V., Wardle,
J. F. C., Roberts, D. H., 1994, ApJ, 435, 140

\noindent Gabuzda D.C., Cawthorne, T. V., Roberts, D. H., Wardle,
J. F. C., 1992, ApJ, 388, 40

\noindent G\'orski K.M., Hivon E., Banday A.J., Wandelt B.D., Hansen
F.K., Reinecke M., Bartelmann M., 2005, ApJ, 622, 759

\noindent Helmboldt, J. F., Taylor, G. B., Tremblay, S., Fassnacht,
C. D., Walker, R. C., Myers, S. T., Sjouwerman, L. O., Pearson, T. J.,
Readhead, A. C. S., Weintraub, L., Gehrels, N., Romani, R. W., Healey,
S., Michelson, P. F., Blandford, R. G., Cotter, G. preprint
(astro-ph/0611459) 

\noindent Henstock, D. R., Browne, I. W. A., Wilkinson, P. N., Taylor,
G. B., Vermeulen, R. C., Pearson, T. J., Readhead, A. C. S. 1995,
ApJS, 100, 1

\noindent Hutsem\'ekers D., 1998, A\&A, 332, 410.

\noindent Hutsem\'ekers D., Lamy H., 2001, A\&A, 367, 381.

\noindent Hutsem\'ekers D., Cabanac R., Lamy H., Sluse D., 2005, A\&A,
441, 915. 

\noindent Jackson N., Battye R. A., Browne I. W. B., Joshi S., Muxlow
T. W. B., Wilkinson P. N., 2007, MNRAS, 376, 371.

\noindent Jain B., Scranton R., Sheth Ravi K., 2003, MNRAS, 345, 62.

\noindent Kendall D.. Young G., 1984, MNRAS, 207, 63.

\noindent Kovalev, Y. Y., Kellermann, K. I., Lister, M. L., Homan,
D. C., Vermeulen, R. C., Cohen, M. H., Ros, E., Kadler, M., Lobanov,
A. P., Zensus, J. A., Kardashev, N. S., Gurvits, L. I., Aller, M. F.,
Aller, H. D. 2005,  AJ, 130, 2473 

\noindent Kovalev, Y. Y., Petrov, L., Fomalont, E. B., Gordon, D.,
2007, AJ, 133, 1236 

\noindent Land K., Magueijo J., 2005, Phys. Rev. Lett., 95, 071301

\noindent Lister M.L., Smith, P.S. 2005, ApJ, 541, 66.

\noindent Lorado T. J., Flanagan E. E., Wasserman I. M., 1997,
Phys. Rev. D, 56, 7507

\noindent Morales J.A, S\'aez, D, 2007, preprint (astro-ph/0701914)

\noindent Myers S.T., Jackson N.J., Browne I.W.A., de Bruyn A.G.,
Pearson T.J., Readhead A.C.S., Wilkinson P.N., Biggs A.D., Blandford
R.D., Fassnacht C.D., 2003. MNRAS 341, 1.

\noindent Nodland B., and Ralston J.P., 1997, Phys. Rev. Lett., 78,
3043 

\noindent Patnaik, A. R., Browne, I. W. A., Wilkinson, P. N.,
Wrobel, J. M., 1992, MNRAS, 254, 655

\noindent Phinney E., Webster R., 1983, Nat, 301, 735.

\noindent Petrov, L., Kovalev, Y. Y., Fomalont, E., Gordon, D., 2005,
AJ, 129, 1163

\noindent Petrov, L., Kovalev, Y. Y., Fomalont, E., Gordon, D., 2006,
AJ, 131, 1872  

\noindent Polatidis, A. G., Wilkinson, P. N., Xu, W., Readhead,
A. C. S., Pearson, T. J., Taylor, G. B., Vermeulen, R. C., 1995, ApJS,
98, 1 

\noindent Pollack, L. K., Taylor, G. B., Zavala, R. T., 2002, AAS,
201, 4817

\noindent Raeth C., Schuecker P., Banday A.J., 2007, preprint
(astro-ph/0702163) 

\noindent Rudnick L., Jones T.W., 1983,  AJ, 88, 518.

\noindent Schwarz D.J., Starkman G.D., Huterer D., Copi C.J., 2004,
Phys. Rev. Lett., 93, 1301S

\noindent Stockman H.S., Moore R.L., Angel J.R.P., 1984, ApJ, 279, 485.

\noindent Taylor, G. B., Vermeulen, R. C., Pearson, T. J., Readhead,
A. C. S., Henstock, D. R., Browne, I. W. A., Wilkinson, P. N. 1994,
ApJS, 95, 345

\noindent Thakkar, D. D., Xu, W., Readhead, A. C. S., Pearson, T. J.,
Taylor, G. B., Vermeulen, R. C., Polatidis, A. G., Wilkinson, P. N.,
1995, ApJS, 98, 33

\noindent V\'eron-Cetty M.-P., V\'eron, P., A Catalogue of Quasars and
Active Nuclei, 2007, 10th edition.

\noindent Visvanathan N., Wills B.J., 1998, AJ, 116, 2119.

\noindent Wardle, J. F. C., Perley, R. A., Cohen, M. H., 1997,
Phys. Rev. Lett., 79, 1801 

\noindent Wills, B.J., Wills D., Breger M Antonucci, R.R.J., Barvainis
R., 1992, ApJ, 398, 454. 

\noindent Wilkinson, P. N., Browne, I. W. A., Patnaik, A. R., Wrobel,
J. M., Sorathia, B., 1998, MNRAS, 300, 790

\noindent Xu, W., Readhead, A. C. S., Pearson, T. J., Polatidis,
A. G., Wilkinson, P. N., 1995, ApJS, 99, 297 

\noindent Zavala, R.T, Taylor, G.B., 2004. ApJ, 612, 749

\noindent Zensus, J.A.,Ros, E., Kellermann,K.I., Cohen M.H.,
Vermeulen,R.C., Kadler, M, 2002, AJ, 124, 662 

\section*{Appendix} {\label{app1}}

\begin{table*}

\begin{tabular}{llrrrrrrr}

\hline

RA&Dec&&&&&&&\\
 HH MM SS&DD MM SS&Name&I(8.4)&p\%(8.4)&p\%(op)&PA(8.4)&PA(op)&PA diff\\

\hline

00 27 15.374 & 22 41 58.17 &  & 320 & 0.378 & 0.63 & 151.3 & 90 & 61.3 \\
01 08 38.771 & 01 35 00.32 &  & 2370 & 2.538 & 1.87 & 110.3 & 143 & -32.7 \\
01 21 56.862 & 04 22 24.75 &  & 1581 & 1.139 & 4.20 & 33.3 & 59 & -25.7 \\
01 26 42.791 & 25 59 01.28 &  & 869 & 2.743 & 1.63 & 155.7 & 140 & 15.7 \\
03 39 30.938 & -1 46 35.81 &  & 2801 & 2.894 & 19.40 & 121.0 & 22 & 81 \\
04 23 15.801 & -1 20 33.06 &  & 4383 & 1.959 & 11.90 & 99.6 & 115 & -15.4 \\
08 08 39.667 & 49 50 36.53 &  & 884 & 1.956 & 8.60 & 108.0 & 179 & -71 \\
08 41 24.366 & 70 53 42.17 &  & 1786 & 6.650 & 1.10 & 106.5 & 102 & 4.5 \\
08 42 05.094 & 18 35 40.99 &  & 856 & 0.253 & 1.74 & 132.0 & 100 & 32 \\
08 54 48.875 & 20 06 30.64 & OJ 287 & 3907 & 2.530 & 10.80 & 79.9 & 156 & -76.1 \\
09 27 03.014 & 39 02 20.85 & 4C 39 & 8456 & 2.314 & 0.91 & 130.2 & 102 & 28.2 \\
09 56 49.876 & 25 15 16.05 &  & 1903 & 0.270 & 1.45 & 37.2 & 127 & -89.8 \\
09 57 38.182 & 55 22 57.74 &  & 1447 & 3.141 & 8.68 & 2.2 & 4 & -1.8 \\
09 58 47.245 & 65 33 54.81 &  & 1276 & 9.601 & 19.10 & 159.5 & 170 & -10.5 \\
10 14 00.478 & 19 46 14.40 &  & 18 & 2.299 & 0.67 & 142.7 & 98 & 44.7 \\
10 41 17.163 & 06 10 16.92 &  & 1359 & 0.864 & 0.62 & 109.8 & 149 & -39.2 \\
10 58 29.605 & 01 33 58.82 &  & 3853 & 1.217 & 5.00 & 122.4 & 146 & -23.6 \\
11 31 09.480 & 31 14 05.49 &  & 125 & 0.796 & 0.95 & 97.9 & 172 & -74.1 \\
11 59 31.834 & 29 14 43.83 &  & 1232 & 0.625 & 2.68 & 167.3 & 114 & 53.3 \\
12 02 40.683 & 26 31 38.63 &  & 78 & 1.290 & 0.65 & 74.5 & 177 & 77.5 \\
12 22 22.550 & 04 13 15.78 &  & 1045 & 0.901 & 5.56 & 87.5 & 118 & -30.5 \\
12 24 52.422 & 03 30 50.29 &  & 825 & 1.546 & 2.51 & 58.0 & 98 & -40 \\
12 24 54.461 & 21 22 46.43 &  & 1067 & 4.607 & 1.52 & 168.6 & 167 & 1.6 \\
12 54 38.256 & 11 41 05.90 &  & 633 & 2.754 & 2.51 & 172.5 & 129 & 43.5 \\
13 10 28.664 & 32 20 43.78 &  & 4029 & 2.577 & 12.10 & 11.6 & 68 & -56.4 \\
13 31 08.302 & 30 30 32.07 & 3C 286 & 2156 & 11.083 & 1.29 & 26.2 & 47 & -20.8 \\
13 43 00.180 & 28 44 07.50 &  & 196 & 0.254 & 0.81 & 175.6 & 45 & 49.4 \\
13 49 34.656 & 53 41 17.04 &  & 742 & 1.884 & 1.73 & 78.2 & 161 & -82.8 \\
15 04 24.980 & 10 29 39.20 &  & 1772 & 2.207 & 3.00 & 179.1 & 160 & 19.1 \\
15 24 41.612 & 15 21 21.06 &  & 335 & 4.557 & 7.90 & 90.2 & 32 & 58.2 \\
15 34 52.454 & 01 31 04.21 &  & 930 & 1.417 & 3.50 & 151.7 & 131 & 20.7 \\
15 40 49.492 & 14 47 45.90 &  & 833 & 3.438 & 17.40 & 156.0 & 145 & 11 \\
15 50 35.270 & 05 27 10.46 &  & 1638 & 1.527 & 4.70 & 11.0 & 14 & -3 \\
15 58 55.185 & 33 23 18.61 &  & 106 & 0.645 & 1.31 & 37.7 & 70 & -32.3 \\
16 08 46.204 & 10 29 07.78 &  & 1767 & 1.792 & 2.10 & 54.5 & 134 & -79.5 \\
16 13 41.065 & 34 12 47.91 &  & 3197 & 5.430 & 1.68 & 5.0 & 134 & 51 \\
16 35 15.493 & 38 08 04.50 &  & 2511 & 0.681 & 2.60 & 65.5 & 97 & -31.5 \\
16 38 13.456 & 57 20 23.98 &  & 1355 & 2.031 & 2.40 & 125.7 & 170 & -44.3 \\
16 42 58.810 & 39 48 37.00 & 3C 345 & 5653 & 4.220 & 4.00 & 24.8 & 103 & -78.2 \\
16 42 07.849 & 68 56 39.76 &  & 1254 & 3.933 & 16.60 & 154.0 & 8 & 34 \\
16 57 20.709 & 57 05 53.51 &  & 517 & 2.814 & 1.34 & 157.6 & 51 & 73.4 \\
17 23 20.797 & 34 17 57.99 &  & 213 & 2.290 & 0.74 & 111.6 & 143 & -31.4 \\
17 40 36.979 & 52 11 43.41 &  & 1357 & 1.174 & 3.70 & 21.1 & 172 & 29.1 \\
17 48 32.841 & 70 05 50.77 &  & 573 & 3.404 & 11.50 & 91.8 & 112 & -20.2 \\
21 23 44.518 & 05 35 22.10 &  & 1539 & 2.679 & 10.70 & 64.0 & 68 & -4 \\
21 48 05.459 & 06 57 38.61 &  & 8042 & 0.557 & 0.60 & 70.0 & 138 & -68 \\
22 32 36.409 & 11 43 50.89 & CTA 102 & 3029 & 1.290 & 7.30 & 89.0 & 118 & -29 \\
22 50 25.343 & 14 19 52.03 & 3C 4543 & 591 & 2.854 & 1.39 & 100.8 & 75 & 25.8 \\
22 53 57.748 & 16 08 53.56 &  & 11031 & 2.377 & 2.90 & 5.3 & 144 & 41.3 \\
22 54 09.342 & 24 45 23.47 &  & 456 & 5.593 & 1.34 & 44.1 & 113 & -68.9 \\
22 57 17.564 & 02 43 17.51 &  & 285 & 1.261 & 1.67 & 21.9 & 2 & 19.9 \\
23 04 28.292 & 06 20 08.32 &  & 329 & 2.654 & 3.69 & 108.5 & 163 & -54.5 \\

\end{tabular}

\caption{\small The 52 common objects in JVAS/CLASS surveys and
  Hutsem\'ekers et al (2005) sample. Position information is from
  JVAS/CLASS data. 
  I(8.4) stands for total flux at 8.4 GHz in mJy,
  p\%(8.4) stands for percentage polarized flux at 8.4 GHz,
  p\%(op)stands for percentage polarized flux in optical and PA(op)
  and PA(8.4) stand for polarization position angle in optical and at
  8.4 GHz respectively. PA diff = PA(op)-PA(8.4).
}
\end{table*}

\end{document}